\documentclass[intlimits,twoside,a4paper]{article}
\usepackage[cp1251]{inputenc}
\usepackage[eqsecnum]{cmpj3}
\usepackage{cmpj3}

\issue{2020}{23}{3}{33001}
\doinumber{10.5488/CMP.23.33001}
\title[Noise induced effects at nano-structured thin films growth during deposition in plasma-condensate devices]
{Noise induced effects at nano-structured thin films growth during deposition in plasma-condensate devices}

\author[V.O. Kharchenko, A.V. Dvornichenko, D.O. Kharchenko]{V.O. Kharchenko\refaddr{label1,label2}, A.V. Dvornichenko\refaddr{label2}, 
D.O. Kharchenko\refaddr{label1}}

\addresses{
\addr{label1}Institute of Applied Physics, NAS of Ukraine, 58 Petropavlovskaya St., 40000 Sumy, Ukraine
\addr{label2}Sumy State University, 2 Rimskii-Korsakov St., 40007 Sumy, Ukraine
}
\date{Received March 2, 2020, in final form May 25, 2020}
\begin{document}

\maketitle

\begin{abstract}
We perform a comprehensive study of noise-induced effects in a stochastic model of 
reaction-diffusion type, describing nano-structured thin films growth at condensation. 
We introduce an external flux of adsorbate between neighbour monoatomic layers caused 
by the electrical field presence near substrate in plasma-condensate devices. 
We take into account that the strength of the electric field fluctuates around its mean value.
We discuss a competing influence of the regular and stochastic parts of the external flux 
onto the dynamics of adsorptive system. It will be shown that the introduced fluctuations 
induce first-order phase transition in a homogeneous system, govern the pattern formation 
in a spatially extended system; these parts of the flux control the dynamics of the patterning, spatial order, morphology 
of the surface, growth law of the mean size of adsorbate islands, type and linear size of 
surface structures. The influence of the intensity of fluctuations onto scaling and statistical properties 
of the nano-structured surface is analysed in detail.  This study provides an insight into the
details of noise induced effects at pattern formation processes in anisotropic adsorptive 
systems.
\end{abstract}

\section{Introduction}

Spatial patterns are widely present in different natural dynamical systems.
Their occurrence has been studied for quite a long time with several
applications in different fields, from hydrodynamic systems, plant ecosystems 
to biochemical and neural systems.
The study of patterns can offer useful information on the underlying processes
causing possible changes of the system. Deterministic mechanisms
in pattern formation have been widely studied (see, for example, \cite{poster1}). 
Spontaneous pattern formation and instabilities have been discovered in many physical, chemical and biological systems, such as thin liquid films \cite{NIE1}, sand ripples and dunes \cite{NIE2,NIE3}, 
crystal growth \cite{NIE4}, water waves \cite{NIE5}, electroconvection in liquid crystals \cite{NIE6}.

Nonlinear systems usually exhibit a disordered behaviour in the absence of fluctuations. 
The influence of noise on spatial extended systems has received lots of attention 
\cite{NIE8,NIE9,NIE10,NIE11,NIE14,NIE16,PhysScr11} in the recent decades.
It was shown that noise may give rise to an ordered behaviour and to new dynamical states
\cite{JPS1,JPS2}. In recent decades, many studies have focused on investigations of noise-induced
phenomena which demonstrate a counter-intuitive role for fluctuations leading to self-organization effects. 
Much of the early work dealt with noise-induced phenomena in zero-dimensional systems. More recently, it 
has become widely recognized that the effects of fluctuations on systems with a large number of degrees of freedom, 
the so-called spatially extended systems, play a major role. The most interesting effects in spatially extended systems 
are noise-induced spatial patterns and phase transitions \cite{JPS4,JPS5,Elder,PhysA2010}. 
The ordering phase transition is associated with the ordered phase (in a thermodynamic sense) 
realization, when a randomly fluctuating source is introduced into the dynamical
system \cite{Buceta9,Buceta10,Buceta11,Buceta12}. 
From a fundamental point of view, such effects are of the dynamic origin: in the short-time limit, 
fluctuations destabilize the disordered homogeneous state. 
The noise plays an organizing role if its amplitude depends on the field variable \cite{JPS1,JPS2}. 
Moreover, the ordered phase can exist for a particular range of the  noise intensities. These effects 
are known as reentrant phase transitions, when an increase in the noise intensity leads to the 
formation of an ordered state at a fixed range of the noise intensity.  
The above reentrance appears as a result of the combined effect of the nonlinearity of the system and 
the spatial coupling. 

Among mathematical models used to perform theoretical studies of noise induced effects in spatially 
extended systems one can issue reaction-diffusion systems. These systems play an important role in 
the study of the generic spatio-temporal behaviour of non-equilibrium systems. Various theoretical analyses 
and computer simulations have demonstrated a possibility of spatio-temporal pattern formation in such systems. 
However, this was not experimentally verified until 1990, when the stationary pattern was first observed 
experimentally in the reaction-diffusion process \cite{Castets1990}.
Reaction-diffusion systems are naturally applied in chemistry, biology, geology, ecology and physics
\cite{RDS1,RDS2,RDS3}. They also were exploited in studying the pattern formation at deposition from gaseous 
phase \cite{Mikhailov1,Mikhailov2,PhysScr12}. In such systems, fluctuations are usually assumed to be 
negligibly small and are taken into account as an additive noise. At the same time, it was previously shown  
that internal multiplicative noise satisfying fluctuation-dissipation relation is capable of controlling the dynamics of pattern formation and statistical properties of the spatial structures (see~\cite{PRE12}). 

In this paper we are aimed at performing a detailed study of the external noise influence onto the dynamics of the 
reaction-diffusion system, describing nano-structured thin films growth in plasma-condensate devices.
We  discuss the noise-induced first-order phase transitions and noise-induced pattern formation. 
We  show a possibility of the noise-sustained reentrant pattern formation and transformation in the 
surface morphology with variation in the intensity of the external fluctuations. 
We  provide statistical analysis of the surface patterns and discuss an influence of the noise-over-signal 
ratio onto the scaling dynamics of the mean size growth of separated nanosized adsorbate islands at deposition. 

The work is organized in the following manner. In the next section we derive the stochastic 
model of anisotropic plasma-condensate system by taking into account the stochastic nature of the 
external flux caused by the electrical field presence near the substrate. In  section~3 we  
discuss noise-induced first-order phase transition in a homogeneous system and noise induced effects 
in a spatially extended system in the framework of stability analysis and numerical simulations. 
We conclude in the last section.

\section{Mathematical model}

\begin{figure}[!t]
\centering\includegraphics[width=0.5\textwidth]{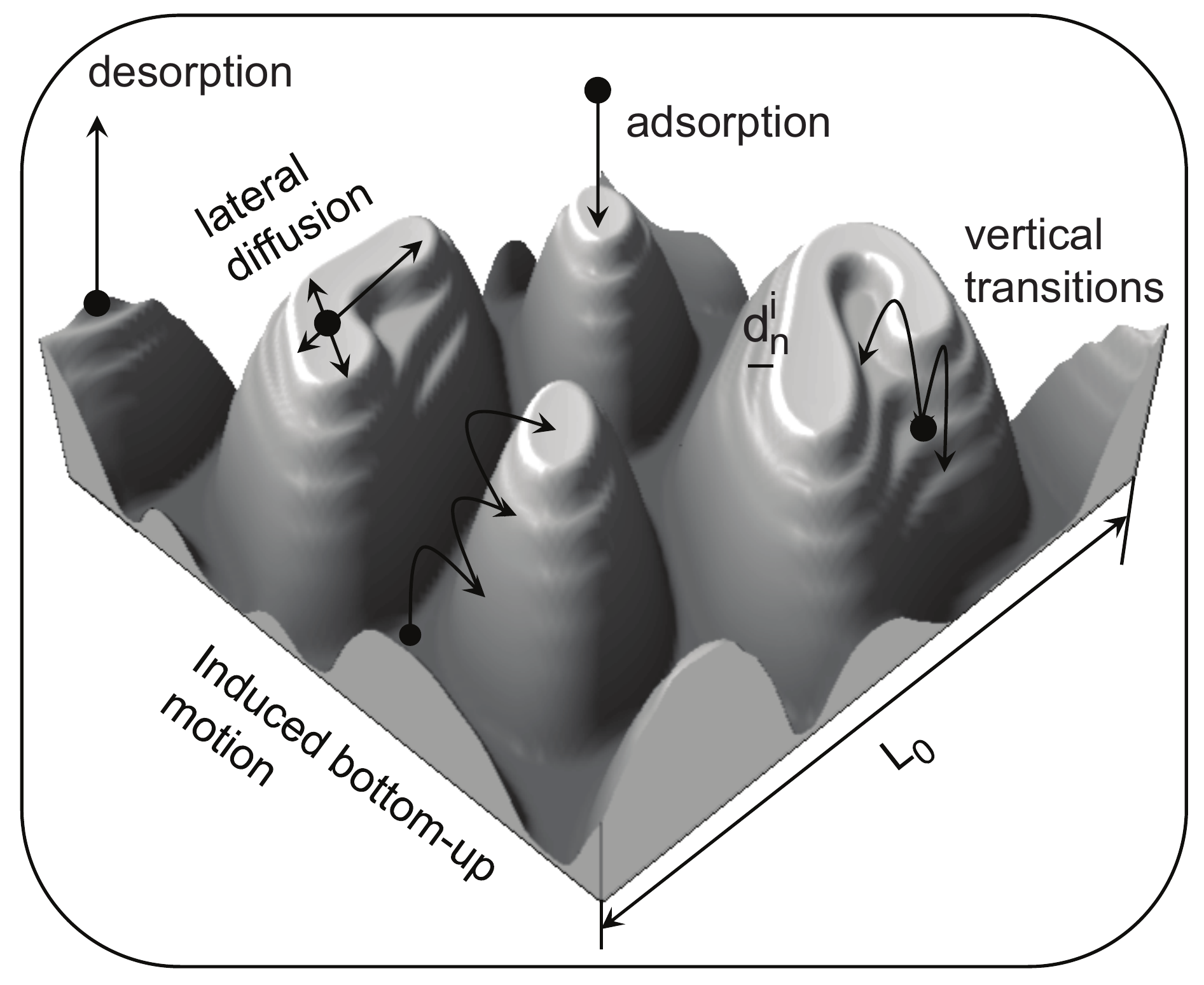}
\caption{Schematic presentation of the basic mechanisms 
describing the deposition of a monoatomic film on a substrate in adsorptive multilayer system.}
\label{fig1}
\end{figure}

During deposition at condensation from gaseous phase or plasma, an evolution of the  local coverage 
of adsorbate on a substrate is caused by the main processes: adsorption, desorption, isotropic 
transference of adatoms between neighbour layers (for multilayer deposition), lateral diffusion of adatoms 
on any adsorbate layer and possible anisotropic motion induced by external fields.
A competition of these mechanisms results in spatio-temporal change in the local adsorbate concentration 
and pattern formation on the growing thin film. The schematic presentation of the basic mechanisms 
describing the deposition of a monoatomic film on a substrate is shown in figure~\ref{fig1}.
In this article we  study the processes of separated surface structures formation at condensation. To this end, we consider  two types of the growth of thin films  namely the Volmer-Weber type of the growth of separated islands  of a thin film when interface energy is relatively large and separated adsorbate islands form and grow before a layer is completed by infilling; and Stranski-Krastanov type when  interface energy is comparable to the island interaction energy, and hence layer formation competes with the formation of islands. During these growth processes, the morphology of the current layer inherits the morphology of the precursor layer as it is shown in figure~\ref{fig1}.

To perform analytical description of the adsorbate concentration evolution in adsorptive multilayer system, 
we  exploit the mesoscopic approach by monitoring the coverage field $x_n({\bf r},t)$ on 
any $n$-th layer; ${\bf r}$ is the space coordinate and $t$ is the deposition time; $n=1\ldots N$, where 
$N$ is the total number of layers. 
The whole layer with the linear size $L_0$ is divided into unit cells with the linear size $\ell$. The local coverage in each unit cell on the $n$-th layer is 
defined as the ratio between the number of adatoms and the number of possible free sites, which yields $x_n\in[0,1]$ with $x_0=1$ being the substrate. In such mesoscale approach, the local distribution of the concentration $x_n$ in each cell is not taken into account.
This approach allows one to construct the reaction-diffusion model 
for adsorbate concentration evolution in each unit cell of a standard form:
\begin{equation}
\partial_t x_n({\bf r},t)=f_n-\nabla\cdot{\bf J}_n,
\label{eq1}
\end{equation}  
where the term $f_n$ is responsible for the quasi-chemical reactions on the $n$-th layer, including adsorption, 
desorption and transfer of adatoms between layers;  the flux ${\bf J}_n$ relates to the mass transport. 

During deposition, an atom (or ion) from gaseous phase (or plasma) can attract the growing 
surface and become adatom with the rate $k_\text{a}$. An adsorption rate $k_\text{a}$ is defined as 
$k_\text{a}=\varpi p\exp(-E_\text{a}/T)$, where $p$ is the pressure inside a chamber, 
$E_\text{a}$ is the activation energy for adsorption, $T$ is the temperature measured here in energetic units, and 
$\varpi$ sets the frequency factor. Adsorption is possible if there are free sites on the current 
$n$-th layer for adsorption; the occupied sites 
on the precursor $(n-1)$-th layer, serving a substrate for adsorption,  and free sites on the next $(n+1)$-th 
layer. Hence, the adsorption processes are described by the term $k_\text{a}x_{n-1}(1-x_n)(1-x_{n+1})$. 
Adsorbed particles (adatoms) can desorb from the $n$-th layer back to the gaseous phase (plasma) with the 
rate $k_\text{d}$, which includes desorption rate for noninteracting particles $k_\text{d}^0=\varpi \exp(-E_\text{d}/T)$, where 
$E_\text{d}$ is the activation energy for desorption, and the contribution caused by the strong local bond 
(substratum-mediated interactions) $\exp(-U_n/T)$, defined by the interaction potential 
$U_n({\bf r})$. The desorption rate $k_\text{d}^0$ relates to the life time scale of adatoms $\tau_\text{d}$  as 
$\tau_\text{d}=[k_\text{d}^0]^{-1}$. For the multilayer system, the desorption processes require occupied sites 
on the precursor $(n-1)$-th layer and free sites on the next $(n+1)$-th layer and are described by the 
term $-k_\text{d}x_nx_{n-1}(1-x_{n+1})$. For the multilayer adsorptive system, one should include into the model the term 
related to the transference of adatoms between neighbour layers, representing standard vertical diffusion 
in the form $w_\updownarrow(x_{n-1}+x_{n+1}-2x_n)$, where $w_\updownarrow$ is the frequency of such transitions, which defines the life time of the adatom on the current layer $\tau_n=[w_\updownarrow]^{-1}$ \cite{CWM2002}. 

The lateral diffusion flux on  any $n$-th layer ${\bf J}_n$ is defined 
as a combination of the free lateral diffusion $-D_\leftrightarrow\nabla x_n$ and diffusion caused by the 
interaction potential $U_n(r)$ in the form $-D_\leftrightarrow/T\mu(x_n)\nabla U_n$; $D_\leftrightarrow$ is the 
lateral diffusion coefficient and $\mu(x_n)=x_n(1-x_n)$ indicates that this diffusion is possible 
on free sites only. By exploiting self-consistence approximation, the interaction potential $U_n(r)$ can be 
defined through the binary attractive potential $u(r)$ for adatoms separated by a distance $r$ in the form
$U_n(r)=-\int u(r-r')x_n(r'){\rm d}r'$
\cite{Wolgraef2003,Wolgraef2004,PhysScr12,Mikhailov1,Mikhailov2,PRE12,SS14,SS15,NRL17,CWM2002}.
In the simplest case, we use $u(r)$ in the Gaussian form
$u(r)=2\epsilon(4\piup r_{0}^2)^{-1/2}\exp\left(-r^2/4r_{0}^2\right)$, where $\epsilon$ and $r_0$ are 
interaction energy and interaction radius, respectively. 
By taking into account that adsorbate concentration 
varies slowly within the interaction radius, we expand the integral $\int u(r-r')x_n(r'){\rm d}r'$:
\begin{equation}
\int u(\mathbf{r}-\mathbf{r}')x_n(\mathbf{r}'){\rm d}\mathbf{r}'\simeq 
\int u(\mathbf{r}-\mathbf{r}')\sum_m \frac{(\mathbf{r}-\mathbf{r}')^m}{m!}\nabla^m
x_n(\mathbf{r}){\rm d}\mathbf{r}'.
\label{int}
\end{equation}
By considering the multilayer system, we assume that these lateral interactions are mediated by the precursor layer with the concentration $x_{n-1}(r)$. By substituting $u(r)$ into equation~(\ref{int}) and taking into account that $r_0^{2m}\to0$ at $m>2$ 
we get an expression for the interaction potential $U_n(r)$ for the $n$-th layer in the following form \cite{SS15,NRL17,CWM2002}:
\begin{equation}
U_n(r)\simeq-\epsilon x_{n-1}[x_n+(1+r_0^2\nabla^2)^2x_n],
\label{eq2}
\end{equation}
where the multiplier $x_{n-1}$ denotes the interactions mediated by the precursor layer; 
$\int u(r)x_n(r){\rm d}r=2\epsilon x_n$, $1/2!\int u(r)r^2\nabla^2x_n(r){\rm d}r=2\epsilon r_0^2\nabla^2x_n$, 
$1/4!\int u(r)r^4\nabla^4x_n(r){\rm d}r=\epsilon r_0^4\nabla^4x_n$.
If the lattice misfit between film and substrate is large then the bonding between adatoms on the first 
layer and substrate is strong \cite{Wolgraef2003}. 
In such a case, the interaction potential (\ref{eq2}) should be generalized 
by taking into account the elastic effects coming from attractive interaction  between the substrate and 
adsorbed particles in the form $U_1^{\text{el}}=\epsilon_s<0$ \cite{Wolgraef2003,Wolgraef2004}. In such 
systems, both elasticity and stress effects make desorption negligible during the growth of the first layer. 
For sufficiently small lattice misfit between film and substrate, elasticity and stress effects even for the 
first growing layer may be neglected \cite{Walg18}. 
If the deposited atoms are of the same type, then on the $n$-th layer with $n>1$ there are no 
elastic effects. They affect onto spatial rearrangement of adatoms on the first layer only. 
We assume that spatial 
configuration of adsorbate on any $n$-th layer with $n>1$ repeats the configuration on the precursor layer. The corresponding explanations were given by independent modelling of multilayer growth discussed in \cite{SS15,NRL17,CrysGrowth19}.
Therefore, next, without loss of generality we neglect the elastic effects.

In special kind of devices used for fabrication of nano-structured thin films, one operates with external fields to 
produce surface patterns of a different type. The typical example is the effect of electromigration, when
ionic transport in the reverse direction of an electrical field is caused by the momentum transfer from free 
electrons to metal ions. This method is used to fabricate line-type structures \cite{JAC,spectr}. 
In accumulative ion-plasma devices, the patterning of adsorbate 
is sustained by the electrical field near the substrate. In this kind of systems, the adsorbed particles can 
desorb back into plasma to be additionally ionized and adsorbed onto higher levels of the adsorbate 
structure of multi-layers \cite{Perekrestov1,Perekrestov2}. In order to describe this induced vertical 
motion of adatoms from the lower layers towards the upper ones, we introduce an additional contribution 
$D_E[x_{n-1}(1-x_n)-x_n(1-x_{n+1})]$, where 
$D_E=|{\bf E}|Ze/T$ is proportional to the strength of the electrical field near the substrate $|{\bf E}|$; 
$Z$ is the coordination number and $e$ is the electron charge \cite{NRL17,CrysGrowth19}.

To describe an evolution of the growing surface in more realistic conditions, one should take into account  the
stochastic nature of the electrical field. It means that the strength of the electric field $|{\bf E}|$ can be 
considered as a fluctuating parameter of the model.  Considering small deviations from the fixed strength 
$|{\bf E_0}|$ we can expand the reaction term $f_n(D_E)$ in equation~(\ref{eq1})  in the vicinity of 
$D_E^0=|{\bf E_0}|Ze/T$, 
which  yields: $f_n=f_n(D_E^0)+\left.\left(\partial f_n\partial D_E\right)\right|_{D_E=D_E^0}\xi$, 
where $\xi$ is assumed to be the stochastic field, $\xi=\xi({\bf r},t)$, that  in the simplest case represents 
zero-mean  white Gaussian noise with correlation 
$\langle\xi({\bf r},t)\xi({\bf r'},t')\rangle=2\sigma^2\delta({\bf r}-{\bf r'})\delta(t-t')$,
where $\sigma^2$ is the intensity of fluctuations of the electrical field strength, proportional to $|{\bf E_0}|$. 
 
In \cite{CWM2002} by studying the 
two-layer model, the authors showed that in a multi-layer system the occupation of each layer affects the one in 
the immediate  neighbor layers. In order to perform numerics for the $N$-layer system, 
one needs to solve $N$ differential equations (\ref{eq1}). In order to characterize 
the influence of the introduced fluctuations  onto the dynamics of pattern formation in the studied system, 
we  pass to the effective 1-layer model, by considering the spatio-temporal evolution  
of adsorbate concentration on the intermediate $n$-th layer. To that end, we  define 
the adsorbate concentration on both precursor $(n-1)$-th and the next $(n+1)$-th layers through one on 
the current $n$-th layer. Let us consider the mean adsorbate concentration 
on  any $n$-th layer as the ratio between square covered by adsorbate on the $n$-th layer $S_n$ and 
square of the substrate $S_0$, as $\langle x_n\rangle=S_n/S_0$:
$S_0\propto L_0^2$, where $L_0$ is the linear size of the substrate;
$S_n=\sum_i ^{M}s_{ni}=\piup\sum_i^{M} r_{ni}^2$, where $s_{ni}$ is the square of the $i$-th structure 
on the $n$-th layer; sum is taken over all $M$ structures. 
Following the  principle of minimization of the surface energy, 
we take into account that the linear size of  each $i$-th multi-layer structure decreases with the layer number $n$
growth by the terrace width $d=\overline{\langle d^i_n\rangle}$, averaged over all $M$ structures and over  
all $N$ layers (see figure~\ref{fig1}).
Formally, we can combine all $M$ areas covered by the adsorbate on the $n$-th layer into one structure
with the linear size $r_n$, which yields $S_n\propto r_n^2$. In such a case, $r_n$ decreases with the layer number 
$n$ by the value $\Delta$, representing the terrace width for the constructed multi-layers structure. 
From a naive consideration it follows that $d<\Delta<L_0$ and $\Delta=\Delta(d,L_0,N)$. In the simplest case, we can put $r_n=r_1-(n-1)\Delta$.
This expression gives the relation between the area covered by adsorbate on  any $n$-th and first layers:
$S_n\propto S_1[1-(n-1)\Delta/r_1]^2$. Hence, for the mean adsorbate concentration on the $n$-th layer,
nearest $(n-1)$-th and $(n+1)$-th layers, we get:  
\begin{eqnarray}
\langle x_n\rangle&=&\frac{S_1}{S_0}\left[1-(n-1)\frac{\Delta}{r_1}\right]^2,\nonumber\\
\langle x_{n\pm1}\rangle&=&\frac{S_1}{S_0}\left\{\left[1-(n-1)\frac{\Delta}{r_1}\right]\mp\left[\frac{\Delta}{r_1}\right]\right\}^2\nonumber\\
&=&\underbrace{\frac{S_1}{S_0}\left[1-(n-1)\frac{\Delta}{r_1}\right]^2}_{\langle x_n\rangle}\mp
2\frac{\Delta}{r_1}\frac{S_1}{S_0}\underbrace{\left[1-(n-1)\frac{\Delta}{r_1}\right]}_{\sqrt{\langle x_n\rangle}\sqrt{S_0/S_1}}
			+\frac{S_1}{S_0}\left(\frac{\Delta}{r_1}\right)^2.		
\end{eqnarray}
After a simple algebra for the averaged concentrations on the precursor layer one finds: $\langle x_{n\pm1}\rangle=\left(\sqrt{\langle x_n\rangle}\mp\beta/2\right)^2$, where $\beta=2\Delta/L_0$. From the naive consideration it follows that for the formation of adsorbate islands, the relation $\Delta<L_0/2$ should be satisfied. This gives $\beta<1$. Physically, the terrace widths $\Delta$ and $d$ depend on the material properties, temperature and deposition conditions and can be defined in real experiments \cite{terrace1,terrace2,terrace3,terrace4}.  
Finally, by taking into account that the morphology of any ($n+1$)-th layer inherits the morphology of the $n$-th layer the same relation is realized with the spatial distribution of the adsorbate concentration. This allows us to use, for the local concentration of adsorbate in each unit cell on the neighbor ($n+1$)-th and ($n-1$)-th layers, the following relations: 
\begin{equation}
x_{n\pm1}(r)=\left(\sqrt{x_n(r)}\mp\beta/2\right)^2.
\label{eq3}
\end{equation} 

Next, it is more convenient to scale time in dimensionless units 
$t/k_\text{d}^0$, and introduce dimensionless parameters $\varepsilon\equiv\epsilon/T$, 
$\alpha\equiv k_\text{a}/k_\text{d}^0$, $u_E=D_E^0/k_\text{d}^0$, $D_0\equiv w_\updownarrow/k_\text{d}^0$.
By introducing the diffusion length $L_\text{d}\equiv\sqrt{D_\leftrightarrow/k_\text{d}^0}>r_0$, the evolution 
equation for the adsorbate concentration $x=x_n$ on the intermediate layer of the multi-layer plasma-condensate 
system in the Stratonovich interpretation reads:
\begin{equation}
\frac{\partial x}{\partial t}=f(x)-L_\text{d}^2\nabla\cdot{\bf J} 
+\sigma^2g(x)\frac{{\rm d}g(x)}{{\rm d}x}+g(x)\xi({\bf r},t),
\label{eq7}
\end{equation}  
where  the reaction term $f(x)$ becomes of the form:
\begin{eqnarray}
 f(x)&=&\alpha(1-x)\nu(x)-x\nu(x)\re^{-2\varepsilon x(\sqrt{x}+\frac{1}{2}\beta)^2}\nonumber\\ 
 &+&u_E\beta_0\sqrt{x}(1-2x)+\frac{1}{4}\beta^2(u_E+2D_0),
\label{eq4}
\end{eqnarray}
with $\nu(x)=(\sqrt{x}+1/2\beta)^2\left[1-(\sqrt{x}-1/2\beta)^2\right]$ and 
$g(x)=\left\lbrace (1-x)(\sqrt{x}+1/2\beta)^2-x[1-(\sqrt{x}-1/2\beta)^2]\right\rbrace $. 
The total lateral adsorbate flux ${\bf J}$ reads:
\begin{equation}
{\bf J}=-\left[ \nabla x-\varepsilon\gamma(x)\nabla\left\{x+(1+r_0^2\nabla^2)^2x\right\}\right],
\label{eq5}
\end{equation} 
where $\gamma(x)=\mu(x)(\sqrt{x}+1/2\beta)^2$. In the further study we assume that the time scales $\tau_\text{d}=[k_\text{d}^0]^{-1}$ and $\tau_n=[w_\updownarrow]^{-1}$ are approximately the same, which yields $D_0\simeq1.0$. For the parameter $\beta$ we put $\beta=0.1$ 
meaning that $\Delta=0.05L_0$ for the terrace width of the pyramidal structure, combined from all separated pyramidal-like adsorbate islands.

The main goal of this work is to perform a detailed study of the influence of the introduced 
multiplicative external noise onto the dynamics of pattern formation in the system studied  and statistical 
properties of the surface morphology. 

\section{Results and discussions}

By considering the homogeneous system equation~(\ref{eq7}) with $\nabla\cdot{\bf J}=0$, we  pay most 
of our attention to the study of
the noise induced phase transitions of the first order. Next, in the framework of stability analysis, 
we discuss the possibility of introducing noise to induce ordering/disordering in a 
spatially extended system. Finally, by exploiting numerical simulations we  analyse the influence of 
 fluctuations of the external field onto the dynamics of pattern formation and statistical properties 
of the surface structures formed at deposition. 

\subsection{Noise induced phase transitions}

\begin{figure}[!t]
\centering\includegraphics[width=0.45\textwidth]{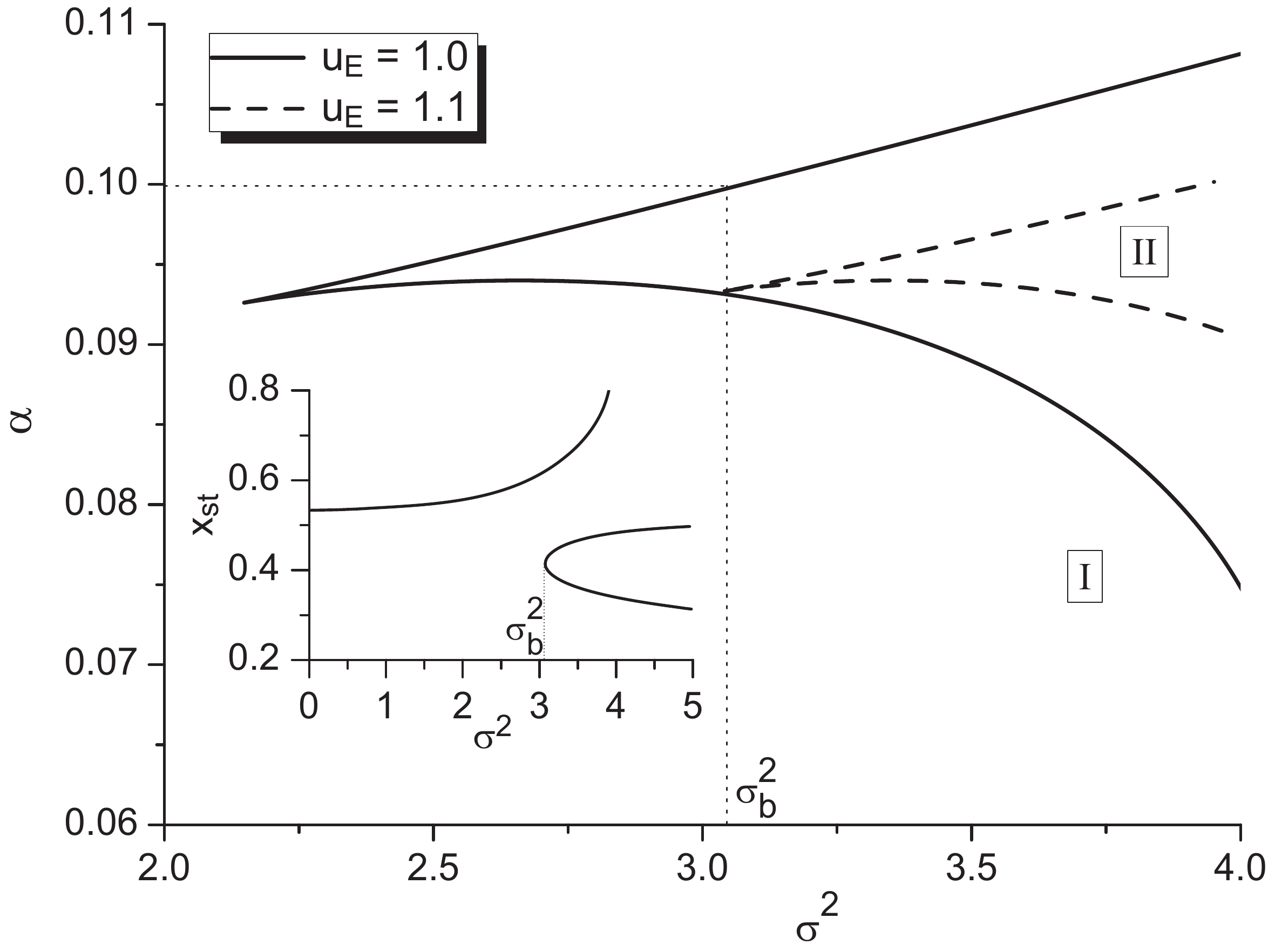}
\caption{Phase diagram of the homogeneous system at $\varepsilon=3.5$ and different values of the anisotropy strength $u_E$. Bifurcation diagram at $u_E=1.0$ and $\alpha=0.1$ is shown in the inset.}
\label{fig2}
\end{figure}

It is known that either internal fluctuations, corresponding to the fluctuation-dissipation relation, or 
external fluctuations can induce phase transitions in complex systems (see, for example 
\cite{VanKampen,JPS1,JPS2}). In this section we  discuss the influence of the introduced 
external fluctuations on stability of the homogeneous state $x_\text{st}$, defined from the equation 
\begin{equation}
f(x)+\sigma^2g(x){\rm d}_xg(x)=0.
\label{f_hom}
\end{equation} 
In the inset in figure~\ref{fig2} we show the bifurcation diagram
of the homogeneous system (\ref{eq7}) at $\varepsilon=3.5$, $u_E=1.0$ and $\alpha=0.1$. 
It follows that with an increase 
in the noise intensity $\sigma^2$, one gets the first order phase transition at $\sigma^2=\sigma^2_b$. 
By varying the adsorption coefficient $\alpha$, we have calculated the critical value of the 
noise intensity responsible for the corresponding bifurcation. 

The phase diagram $\alpha(\sigma^2)$ 
is shown as the main plot in figure~\ref{fig2} for different values of the anisotropy strength $u_E$. 
Here, in the domain I, the system is characterized by the single stationary state $x_\text{st}$; in the 
domain II (inside the corresponding cusp), the system is bistable. It is seen that an increase in the 
anisotropy strength $u_E$ requires elevated values of its fluctuation intensity $\sigma^2$ for the 
first-order transition. Hence, the competition of deterministic and stochastic parts of the external flux, 
induced by the electric field near the substrate, controls phase transitions of the first order in the system. 

In the next sections we  consider a spatially extended system to define the influence of  fluctuations 
of the external flux competing with its deterministic part onto the pattern formation.

\subsection{Stability analysis}

Let us analyse the stability of  homogeneous stationary states $x_\text{st}$ to inhomogeneous 
perturbations in the framework of the standard stability analysis. According to this procedure, 
the deviation of the adsorbate concentration $x$ from the stationary value $x_\text{st}$ is assumed in the 
form $\delta x=x-x_\text{st}\propto \re^{\lambda(k)t}\re^{\ri kr}$, where $k$ is the wave number and 
$\lambda(k)$ is the stability exponent. Assuming $\delta x$ to be a small parameter, we can expand 
the left-hand side of the equation~(\ref{f_hom}) in the vicinity of $x_\text{st}$. In such a case, from the evolution 
equation (\ref{eq7}) one immediately gets the dispersion relation:
\begin{equation}
\lambda(\kappa)
={\rm d}_x\left[f(x)+\sigma^2g(x){\rm d}_xg(x)\right]|_{x=x_\text{st}}
-\kappa^2\left[1-2\varepsilon\gamma(x_\text{st})(1-\rho^2\kappa^2)\right],
\label{eq8}
\end{equation}
where notations $\kappa\equiv kL_\text{d}$, $\rho\equiv r_0/L_\text{d}$ are used and the limit $\rho^4\to0$ is considered. 
The stability exponent $\lambda(\kappa)$ can be associated with the order parameter for spatial instability in 
the  system studied. If  $\lambda(\kappa)<0\ \forall\ \kappa$, then the system is stable, meaning that all spatial 
instabilities will disappear in time. In such a case, the adsorbate will cover the substrate homogeneously 
without any stable separated structures (islands). The existence of  positive values of the stability exponent 
$\lambda(\kappa)$ means that spatial perturbation will grow in time, leading to spatial ordering of the 
coverage field with the formation of separated surface structures. Here, 
one has $\lambda(\kappa)>0$ at $\kappa\in(\kappa_1,\kappa_2)$ and the period of spatial modulations 
$\kappa_{m}$ corresponds to the maximal value of the stability exponent, which can be defined from the 
relation ${\rm d}_\kappa\lambda(\kappa)=0$. 

Next, we refer to the  field that exhibits an ordered state with organized
spatial structures as patterned (i.e., ordered). This general definition, including both periodic and multiscale patterns, is often adopted in the environmental sciences,
where the number of different processes can prevent the organization of
a system with a clear dominant wavelength.

From the stability exponent equation~(\ref{eq8}) it follows that the term with $\kappa^4$ will stabilize the 
system due to $\gamma(x_\text{st})>0$. On the other hand, the first term in the right-hand side in 
equation~(\ref{eq8}) should be negative to enforce the condition $\kappa_1>0$.  Instability of the system 
equation~(\ref{eq7}) is caused by the term $1-2\varepsilon\gamma(x_\text{st})$ with $x_\text{st}=x_\text{st}(\sigma^2)$. 
Hence, one can expect that the variation in the noise intensity  at other fixed parameters can induce 
ordering/disordering of the system. 
 
\begin{figure}[!t]
\centering\includegraphics[width=0.45\textwidth]{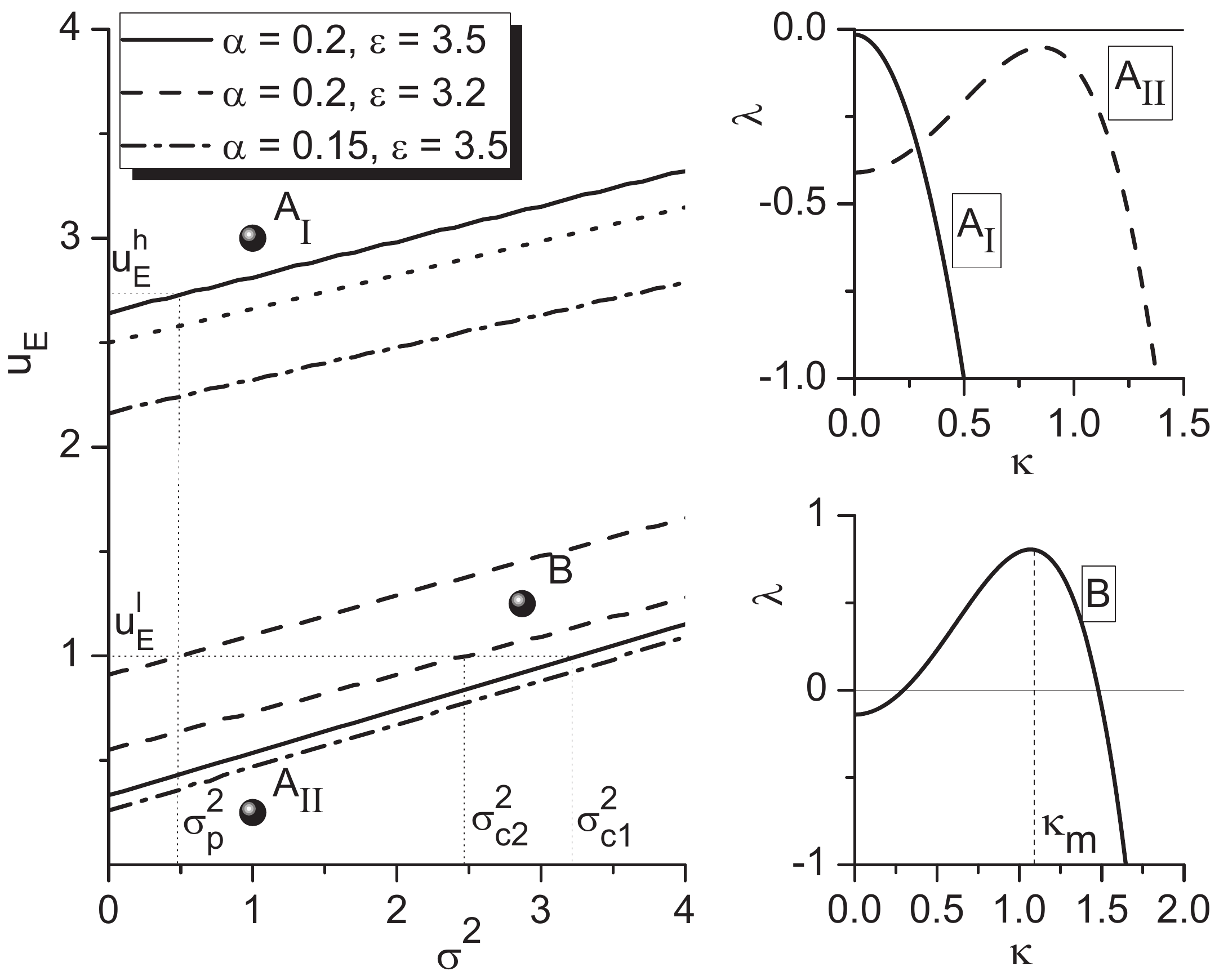}
\caption{Stability diagram $u_E(\sigma^2)$ (in the left-hand panel) and stability exponent $\lambda(\kappa)$ (in the right-hand panel) in different domains of diagram.}
\label{fig3}
\end{figure}

The provided linear stability analysis allows one to define the domains of the main system parameters, 
where the pattern formation is possible. The calculated stability diagram in coordinates 
$(\sigma^2,u_E)$ at different values of $\alpha$ and $\varepsilon$ is shown in the left-hand panel in 
figure~\ref{fig3}.
Here, inside the domains A$_I$ and A$_{II}$, stability 
exponent is negative for all $\kappa>0$ [see typical dependence $\lambda(\kappa)$ in the right-hand panel 
in the top in figure~\ref{fig3}]. In the domain B bounded by curves of the same type 
the stability exponent $\lambda(\kappa)$ becomes positive in the interval $(\kappa_1,\kappa_2)$. 
Typical dependencies of the stability exponent $\lambda$ on the reduced wave-number $\kappa$ 
inside the domain B is shown in the right-hand panel in the bottom in figure~\ref{fig3}.

It follows that in the deterministic quasi-equilibrium 
system with $\sigma^2=0$ and $u\to0$ (weak electrical field near substrate) no spatial instabilities 
can be realized (domain A$_{II}$). Here, only the layer-by-layer growth of the thin film is possible 
without any separated structures \cite{EPJB18}. Let us discuss, initially, the influence of the 
competition of regular and stochastic parts of the external flux, $u_E$ and $\sigma^2$  onto the stability 
of the stationary homogeneous state $x_\text{st}$ to inhomogeneous perturbations at $\alpha=0.2$ and 
$\varepsilon=3.5$ (see solid curves in left-hand panel in figure~\ref{fig3}).  It follows that 
in the case of a strong anisotropy ($u_E=u_E^h$) in quasi deterministic system ($\sigma^2\to0$), 
a rapid bottom-up motion of adatoms leads to a decrease in the adsorbate concentration on the  
layer, and thus the required saturation of the adsorbate concentration needed for patterning is not achieved.
An increase in the noise intensity induces the pattern formation in the system at $\sigma^2=\sigma^2_p$ 
(transition from the domain A$_I$ towards domain B). With a further growth in $\sigma^2$, 
the growing surface remains structured. At intermediate values of the anisotropy strength 
$u_E=u_E^l$ even in a deterministic system, surface structures are stable. Here, an increase in the noise 
intensity $\sigma^2$ provides stabilization of the homogeneous state $x_\text{st}$ at $\sigma^2=\sigma^2_{c1}$. 

A decrease in the interaction strength $\varepsilon$ at a fixed adsorption coefficient leads to a shrink 
in the domain B, where the pattern formation is possible (see dash curves in left-hand panel in figure~\ref{fig3}). 
Here, at fixed values of the deterministic part of the external flux $u_E=u_E^l$ one gets noise induced 
reentrant ordering in the system: in the cases $0\leqslant\sigma^2<\sigma^2_p$ and $\sigma^2>\sigma^2_{c2}$, 
no spatial instability is realized; if $\sigma^2\in(\sigma^2_p,\sigma^2_{c2})$, stable surface structures 
will be formed on the growing surface during deposition. A decrease in the adsorption coefficient 
leads to a decrease in the critical values of the anisotropy strength $u_E$, 
when patterning is possible (compare solid and dash-dot curves in the left-hand panel in figure~\ref{fig3}). 
The dot curve in the left-hand panel in figure~\ref{fig3} corresponds to the relation $u_E=2.5+0.16188\sigma^2$ at 
$\alpha=0.2$, $\varepsilon=3.5$ and will be discussed later.

\begin{figure}
\begin{center}
\includegraphics[width=0.45\textwidth]{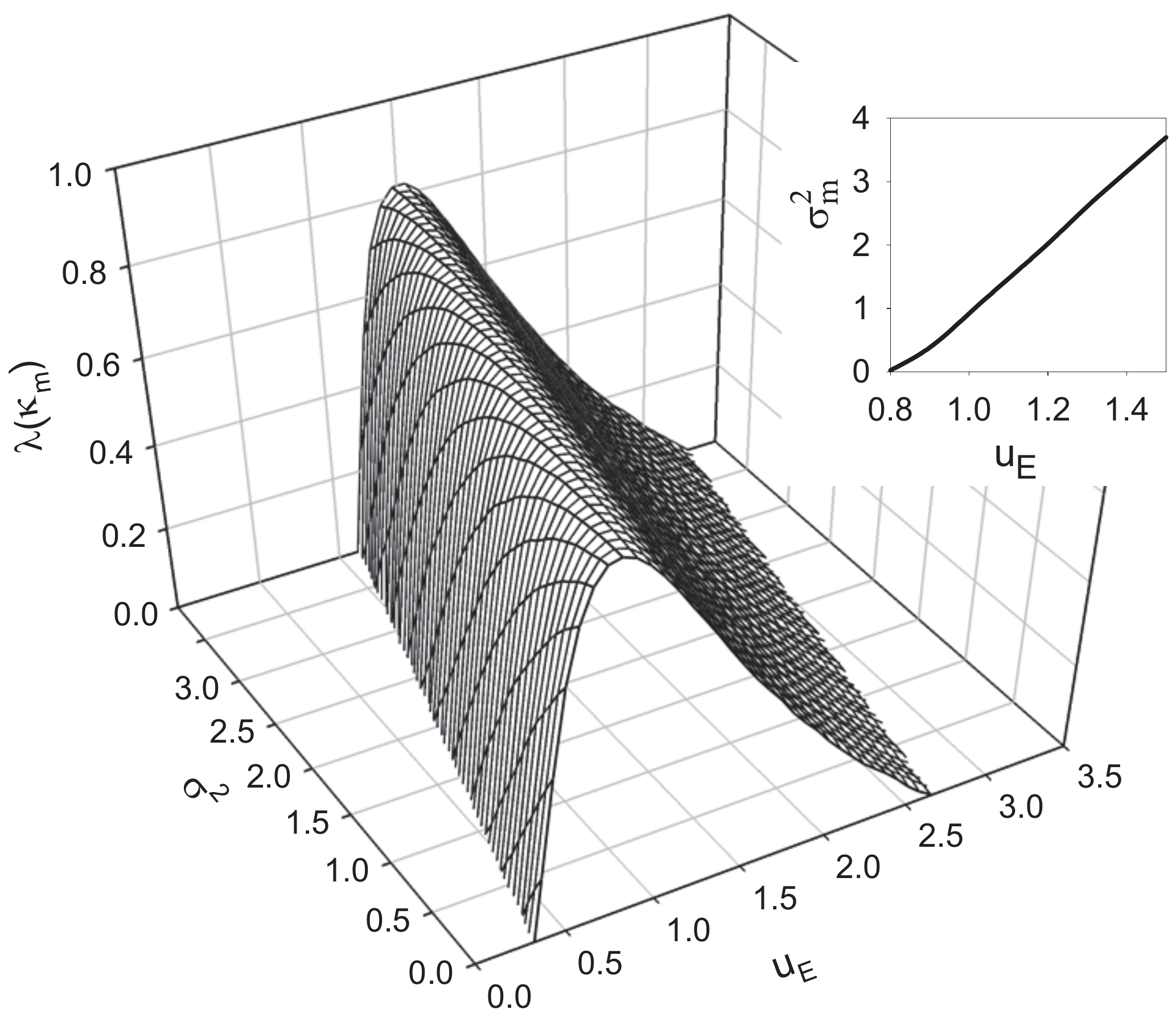}
\end{center}
\caption{Dependencies of the stability exponent $\lambda$, calculated for the most unstable mode $\kappa_m$ at 
$\alpha=0.2$ and $\varepsilon=3.2$.}
\label{fig4}
\end{figure}

Dependence of the maximal value of the stability exponent, corresponding to the most unstable 
mode $\kappa_m$ (see right-hand bottom panel in figure~\ref{fig3}) \emph{versus} anisotropy strength $u_E$ and noise 
intensity $\sigma^2$ at fixed 
$\alpha=0.2$ and $\varepsilon=3.5$ (solid curves in figure~\ref{fig3}) is shown in figure~\ref{fig4}. 
It follows that with an increase in either deterministic or stochastic part of the external flux, the maximal 
value of the stability exponent $\lambda(\kappa_m)$ is characterized by the maximal value. 
It means that the spatial order 
of the system increases, attains the maximal value and then decreases with the growth in either $u_E$ 
or $\sigma^2$. The value $\sigma^2_m$, that corresponds to the maximal value of the stability exponent $\lambda$, increases with $u_E$, which is shown in the inset in figure~\ref{fig4}. Moreover, with an increase 
in the anisotropy strength $u_E$ with $\sigma^2=\sigma^2_m(u_E)$ from the inset in figure~\ref{fig4}, 
the maximal value of the stability exponent $\lambda(\kappa_m)$ increases. Hence, an increase in 
both stochastic and deterministic parts of the external flux promotes the formation of a  well ordered 
surface during deposition in plasma-condensate system. 
Such non-monotonous dependence of the maximal value of the stability exponent on the deterministic and 
stochastic parts of the external flux means a change in the type of the surface structures realized 
during deposition. It was previously shown  that an increase in $u_E$ in a deterministic system 
leads to the change in the surface morphology from separated nano-holes inside an adsorbate matrix 
towards separated multilayer nano-dots  on the substrate \cite{NRL17}. Therefore, one can expect that 
the introduced fluctuations will lead to the morphological transformation of the surface pattern with 
an increase in the noise intensity $\sigma^2$. 

Hence, from the provided stability analysis it follows that 
a competition of the deterministic and stochastic parts of the external flux of adatoms, 
caused by the electrical field near the substrate, leads to the noise induced reentrant scenario of 
pattern formation in an adsorptive plasma-condensate system.

\subsection{Numerical simulations}

To perform a detailed study of an influence of the introduced fluctuations of the external flux 
onto the dynamics of pattern formation and possibility to control both the morphology of the surface 
and the size of islands  in derived stochastic plasma-condensate systems, in this section we 
 perform numerical simulations of the processes of pattern formation during deposition. 
To this end, we solve the Langevin equation equation~(\ref{eq7}) on two-dimensional hexagonal grid with linear size 
$L=256\Delta x$ and periodic boundary conditions. To solve the multiplicative noise Langevin equation 
treated in the Stratonovich sense, we use the Milstein scheme \cite{PRE2010cite31}. 
The white noise source was generated with the help of the Box-Muller algorithm, satisfying 
generation of random numbers with the Gaussian distribution \cite{PRE2010cite32}.
In the case of  triangular or hexagonal symmetry, there 
are three wave vectors separated by $2\piup/3$ angles.
Spatial derivatives of the second and fourth orders were computed according to the 
standard finite-difference scheme for the hexagonal grid. The time step was $\Delta t=10^{-3}$, the spatial integration 
step was $\Delta x=0.5$. As initial conditions we use Gaussian distribution with 
$\langle x({\bf r},0)\rangle=\langle (\delta x({\bf r},0))^2\rangle=10^{-2}$. 
In the computational scheme, the total size of the system is $L\simeq40L_\text{d}$.  

\begin{figure}
\begin{center}
a) \includegraphics[width=0.9\textwidth]{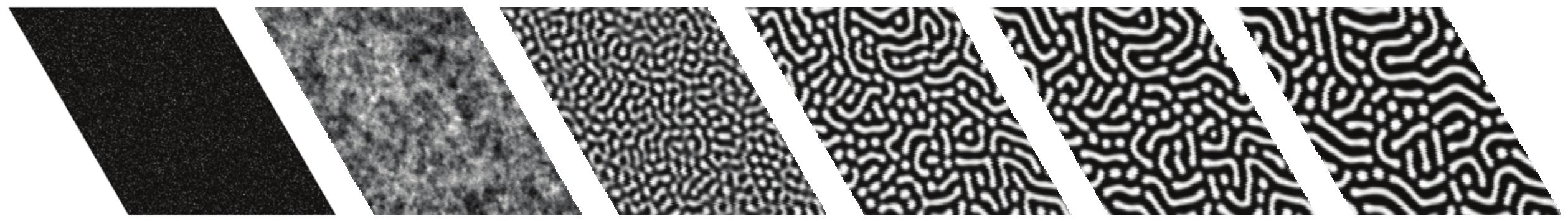}\\
b) \includegraphics[width=0.9\textwidth]{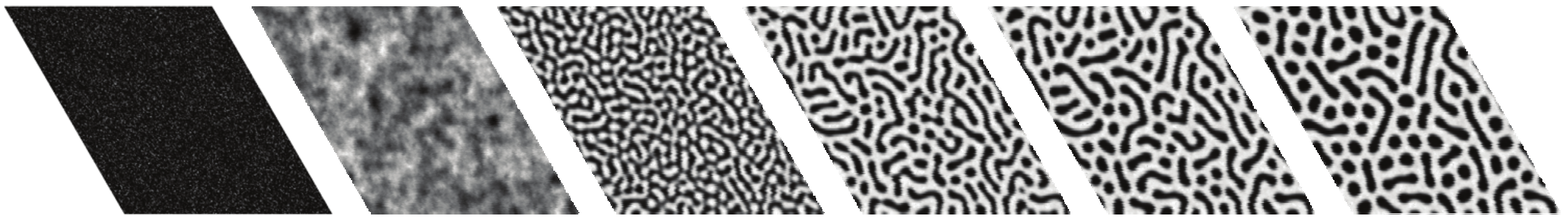}
\end{center}
\caption{Snapshots of the system evolution from left to right 
at $\alpha=0.2$, $\varepsilon=3.5$, $u_E=1.0$ 
and: a) $\sigma^2=0.5$; b) $\sigma^2=2.5$.}
\label{fig5}
\end{figure}

An evolution of the morphology of the growing surface is shown in figures~\ref{fig5} a,b from left to right 
at $\alpha=0.2$, $\varepsilon=3.5$, $u_E=1.0$  and different values of the noise intensity $\sigma^2$: 
$\sigma^2<\sigma^2_m$ and $\sigma^2>\sigma^2_m$, respectively. 
Here, with the help of the shades of the grey color, we show the adsorbate concentration in 
each site of the hexagonal grid: black color corresponds to the spatial configuration without 
adsorbate on the layer; white color means that the current cite is filled with adsorbate. 
It follows that during the deposition after some incubation period an interaction of adsorbate 
leads to pattern formation on the layer: formation of a large amount of small adsorbate clusters. 
These clusters interact and, depending on the deposition conditions, form a stationary picture 
of the surface morphology, shown in the last column in figure~\ref{fig5}. At small values of the noise 
intensity (see figure~\ref{fig5} a), the initially formed small adsorbate clusters grow, while at large times 
one gets a fixed number of adsorbate islands of different forms (spherical and elongated clusters).
At elevated values of the noise intensity, fluctuations result in an increase of the adsorbate concentration 
on the layer, leading to the formation of separated holes of different forms inside the adsorbate matrix 
(see figure~\ref{fig5} b). Hence, an increase of the noise intensity controls the type of the surface patterns 
leading to a morphological transformation of the growing surface from separated adsorbate islands 
on the substrate towards separated holes inside the adsorbate matrix. 

The dynamics of pattern formation in spatially extended systems can be effectively studied 
by monitoring the mean concentration $\langle x \rangle(t)$, averaged over the whole computational grid, 
and the dispersion of the coverage field $\langle(\delta x)^2\rangle(t)=\langle x^2\rangle-\langle x\rangle^2$.
The latter quantity plays the role of an effective order parameter in problems of pattern formation. 
In the case $\langle(\delta x)^2\rangle=0$, there is no essential difference in concentration of adsorbate 
$x$ in different cells of the computational grid. The growing-up temporal dependence 
$\langle(\delta x)^2\rangle(t)$ indicates the ordering processes with the formation of dense (enriched 
by adsorbate) and diluted (depleted by adsorbate) phases: the larger is $\langle(\delta x)^2\rangle$, the 
lager is the order of spatial configuration of adsorbate. If the order parameter $\langle(\delta x)^2\rangle$ 
attains stationary non-zero value, then the spatial configuration becomes stable and no further changes 
in distribution of adsorbate  appear. 

\begin{figure}[!t]
\centering\includegraphics[width=0.49\textwidth]{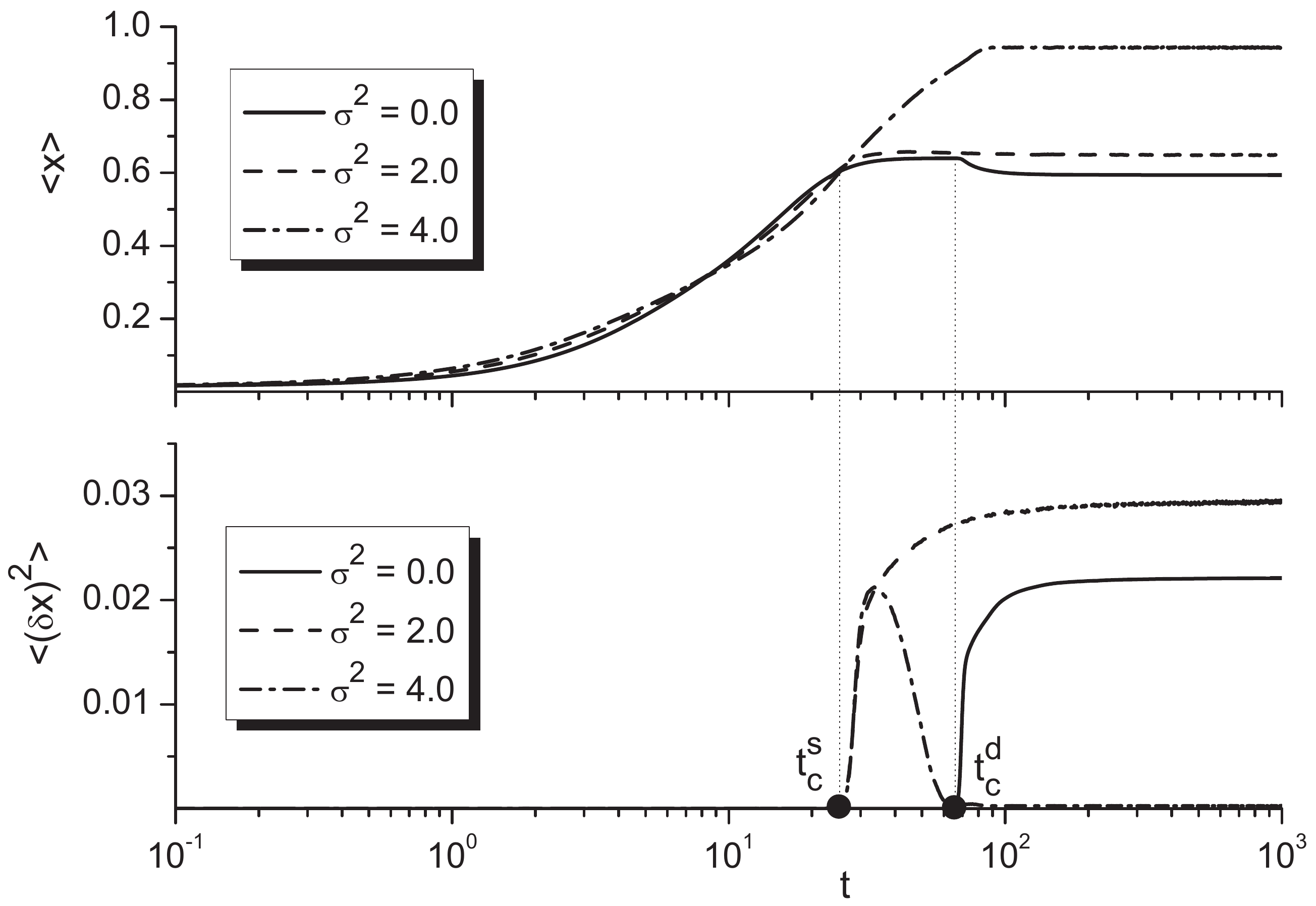}
\caption{Evolution of the mean adsorbate concentration $\langle x \rangle$ at $\alpha=0.2$, $\varepsilon=3.5$, $u_E=1.0$ and different values of the noise intensity $\sigma^2$.}
\label{fig6}
\end{figure}

Temporal dependencies of both the mean adsorbate concentration $\langle x\rangle$ and the order 
parameter $\langle(\delta x)^2\rangle$ are shown in figure~\ref{fig6} at fixed values of the adsorption coefficient 
$\alpha$, interaction strength $\varepsilon$, deterministic part of the external flux (anisotropy strength) 
$u_E$ and different values of the intensity $\sigma^2$ of the fluctuating part of the external field.
Here, the mean concentration in shown in the top panel; dispersion is presented in the bottom panel.
First, let us discuss the reference deterministic case with $\sigma^2=0$, shown by solid curves. 
It follows that during deposition the mean adsorbate concentration grows in time and the dispersion 
of the coverage field equals zero. After some incubation period $t=t_c^d$, the quantity $\langle x\rangle$ 
drops a little and the dispersion starts to grow, meaning rearrangement of adsorbate on the layer 
with the formation of diluted and dense phases. With further deposition, both $\langle x\rangle$ and 
$\langle(\delta x)^2\rangle$ attain their stationary non-zero values $\langle x\rangle_{\text{st}}$ and 
$\langle(\delta x)^2\rangle_{\text{st}}$, respectively, indicating the formation of a well structured 
surface. 

The introduced fluctuations of the external flux influence the dynamics of the mean adsorbate
concentration and its dispersion and their stationary values (see dash curves in figure~\ref{fig6} 
at $\sigma^2=2$). Here, one should indicate a noise induced acceleration of the ordering processes: 
the incubation period $t_c^s$ becomes smaller compared to $t_c^d$ in the deterministic case. Moreover, 
at $t>t_c^s$, there is no  decreasing dynamics of the mean adsorbate concentration. It slowly goes to a 
stationary value which is a little larger than in the deterministic case. Noise action also increases the stationary 
value of the order parameter, leading to the formation of a more ordered film. 
At large values of the noise intensity (see curves at $\sigma^2=4$ in figure~\ref{fig6}), the situation  
changes crucially. Here, noise does not affect the incubation time $t_c^s$, which determines the 
start of the ordering processes (see dash-dot curve in the bottom panel in figure~\ref{fig6}). With a further 
exposing, the mean adsorbate concentration continuously grows in time, whereas the dispersion 
attains a maximal value and then drops towards zero, meaning homogenization of adsorbate 
distribution on the layer. Hence, at large  intensity of fluctuations, only transient patterns are possible. In the 
stationary limit, when the mean concentration does not vary with time, adsorbate with high concentration 
 covers the whole layer and no surface structures can be formed. 

\begin{figure}[!t]
\begin{center}
a) \includegraphics[width=0.45\textwidth]{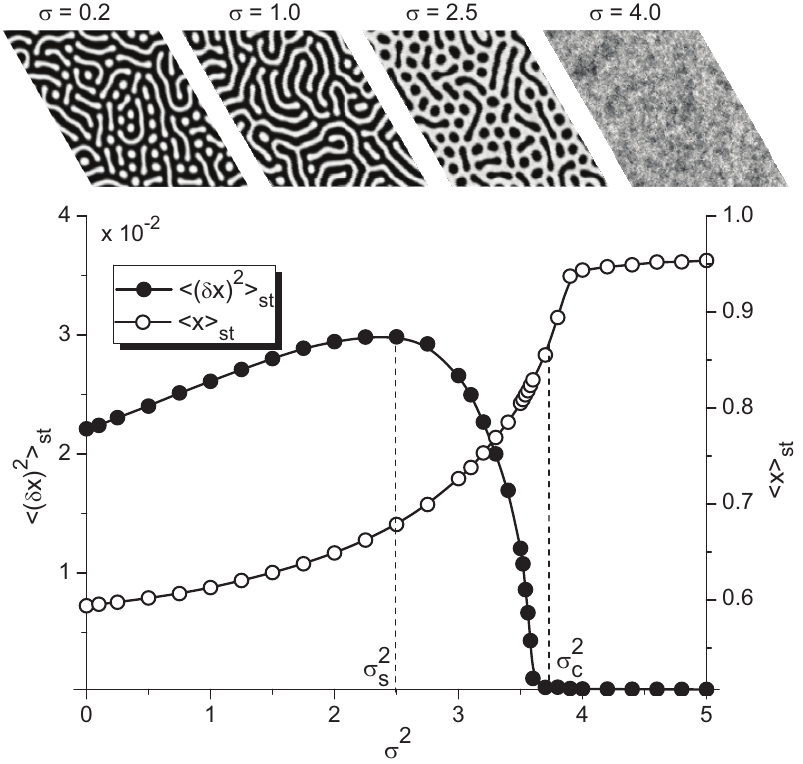}\hfill b)\includegraphics[width=0.45\textwidth]{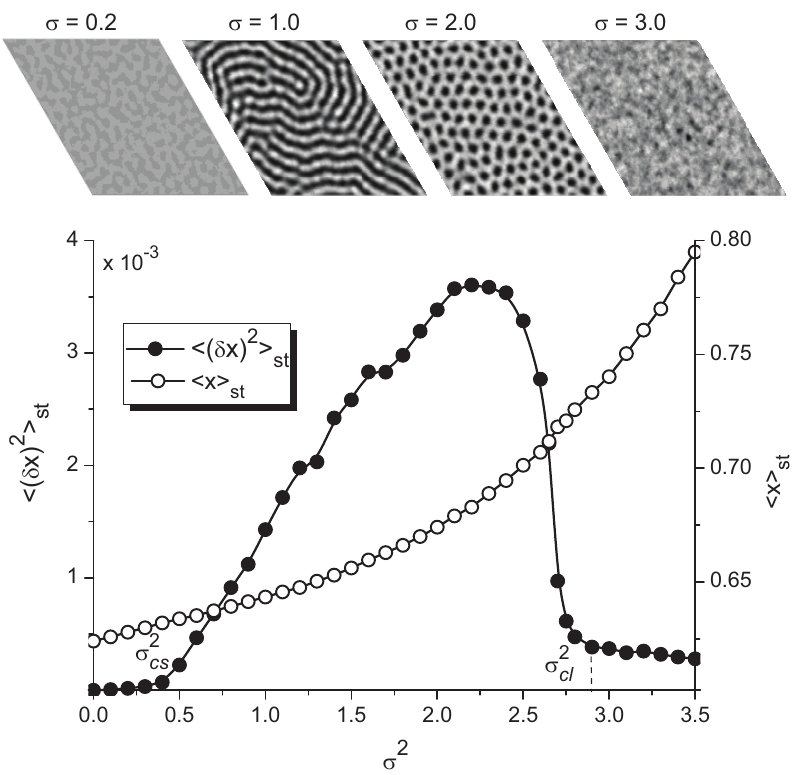}
\end{center}
\caption{Dependencies of the stationary values of adsorbate concentration $\langle x \rangle_\text{st}$ and 
dispersion $\langle (\delta x)^2 \rangle_\text{st}$ on noise intensity $\sigma^2$ at $\alpha=0.2$, $u_E=1.0$ 
and: a) $\varepsilon=3.5$, b) $\varepsilon=3.2$.}
\label{fig7}
\end{figure}

In order to provide a detailed study of the influence of the noise intensity onto stationary picture 
of adsorbate distribution on the layer, next we  analyse the dependencies of both the stationary value 
on the mean adsorbate concentration on the layer $\langle x\rangle_\text{st}$ and the stationary value of the 
order parameter $\langle(\delta x)^2\rangle_\text{st}$ on the noise intensity $\sigma^2$. In figures~\ref{fig7} a, b,
the corresponding dependencies are shown for $\alpha=0.2$ and  different values of the interaction strength 
$\varepsilon$ with the typical quasi-stationary  snapshots at different values of the noise intensity in the top.

Initially, let us discuss the case $\varepsilon=3.5$, shown in figure~\ref{fig7} a. As was shown in 
figures~\ref{fig3} and \ref{fig6}, in such a case, even in deterministic case, one gets a nano-structured thin 
film during deposition. Here, with an increase of the noise intensity $\sigma^2$, both $\langle x\rangle_\text{st}$ 
and the stationary value of the order parameter $\langle(\delta x)^2\rangle_\text{st}$ initially grow. 
At $\sigma^2=\sigma^2_s$ the order parameter attains maximal value and then decreases until 
$\sigma^2<\sigma^2_c$ (see filled circles in figure~\ref{fig7} a). In the case 
$\sigma^2>\sigma^2_c$, one gets $\langle(\delta x)^2\rangle_\text{st}\simeq0$, whereas the stationary value 
of the adsorbate concentration continues to grow (empty circles in figure~\ref{fig7} b). 
At large values of the noise intensity, its growth does not affect the quantity $\langle x\rangle_\text{st}$. 
In the top panel in figure~\ref{fig7}, a  typical snapshots for the 
stationary picture of the spatial adsorbate distribution on the layer show a change in the surface morphology 
with an increase  of the noise intensity $\sigma^2$. By comparing the results of the stability analysis and numerical
simulations, one has $\sigma^2_m<\sigma^2_s$ (see insertion in figure~\ref{fig4}). It means that the maximal value 
of the stability exponent $\lambda(\kappa_m)$, that corresponds to the $\sigma^2_m$, relates with critical value 
of the noise intensity when the morphological transformation of the surface is realized, and at 
$\sigma^2=\sigma^2_m$, one gets labyrinthine-like structure with percolating clusters of adsorbate.
The maximal spatial order is observed at $\sigma^2=\sigma^2_s$, when separated holes are formed inside the
adsorbate matrix (see snapshot at $\sigma^2=2.5$ in the top panel in figure~\ref{fig7}~a).

Next, let us consider the case of small interaction strength $\varepsilon=3.2$, shown in figure~\ref{fig7} b. 
In such a case, the stationary value of the adsorbate concentration continuously grows with the noise intensity 
(see empty circles in figure~\ref{fig7} b). The dependence of the stationary value of the order parameter on 
the noise intensity, discussed in the previous section,  shows a reentrant picture of ordering. In the 
case of deterministic system and weak fluctuations, $\sigma^2<\sigma^2_{cs}$, one has $\langle(\delta x)^2\rangle_\text{st}\simeq0$, meaning homogeneous distribution of adsorbate over the layer without any 
spatial structures (see snapshot at $\sigma^2=0.2$ in the top in figure~\ref{fig7} b). At 
$\sigma^2_{cs}<\sigma^2<\sigma^2_{cl}$, the stationary value of the order parameter increases with the 
noise intensity growth, attains maximal value and decreases; the morphology of the surface transforms from separated adsorbate islands towards separated holes (see filled circles in figure~\ref{fig7}~b and typical 
snapshots in the top panel). At elevated values of the noise intensity ($\sigma^2>\sigma^2_{cl}$), noise acts 
in the same manner as in the case of large values of the interaction strength leading to the 
homogenisation of the adsorbate distribution on the layer (see snapshot at $\sigma^2=3.0$ in figure~\ref{fig7} b). 
Hence, the performed numerics confirm the results of the stability analysis regarding the noise induced reentrant
ordering in the  system studied. 

\begin{figure}[!t]
\begin{center}
a) \includegraphics[width=0.22\textwidth]{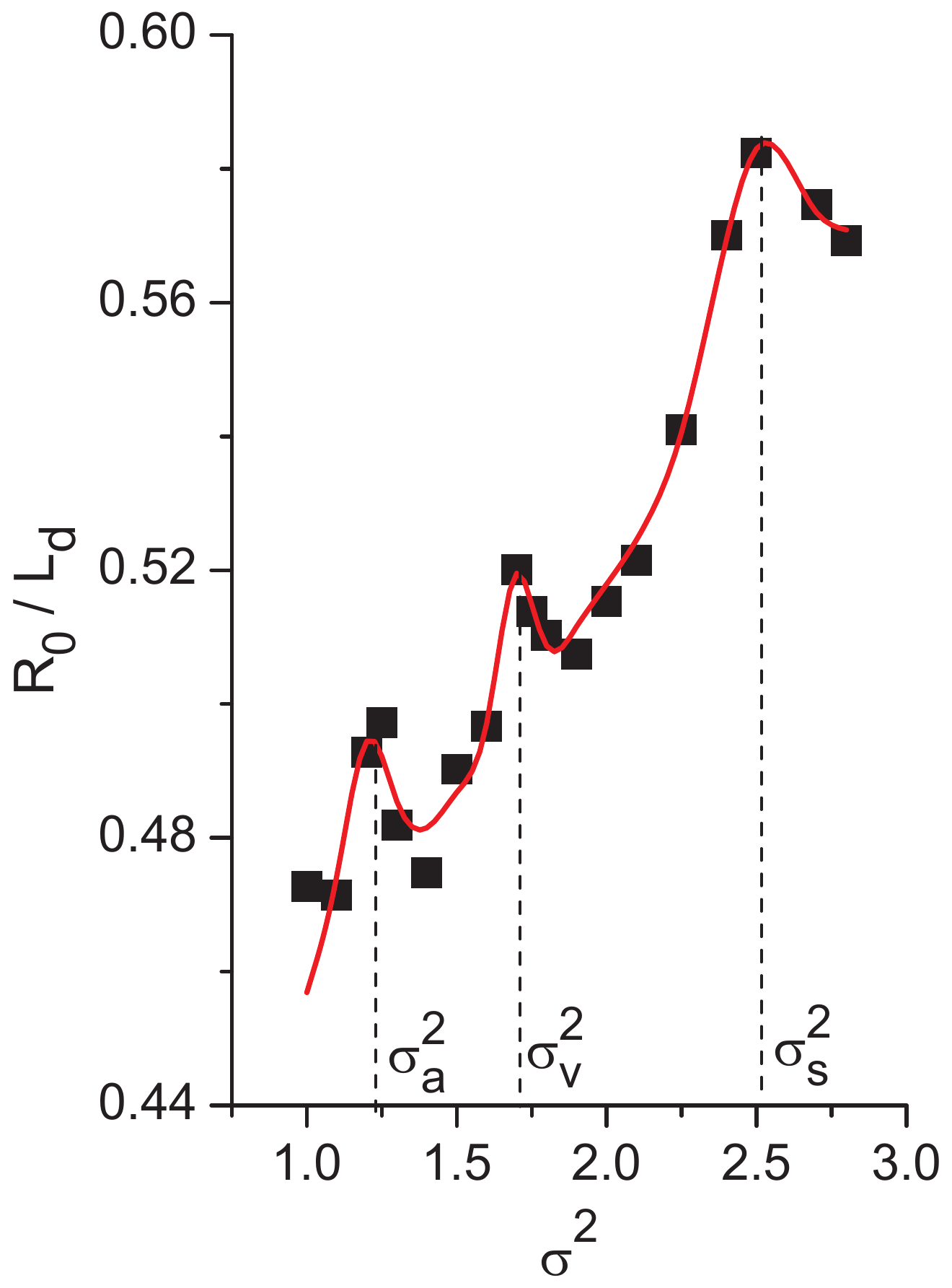}
b) \includegraphics[width=0.22\textwidth]{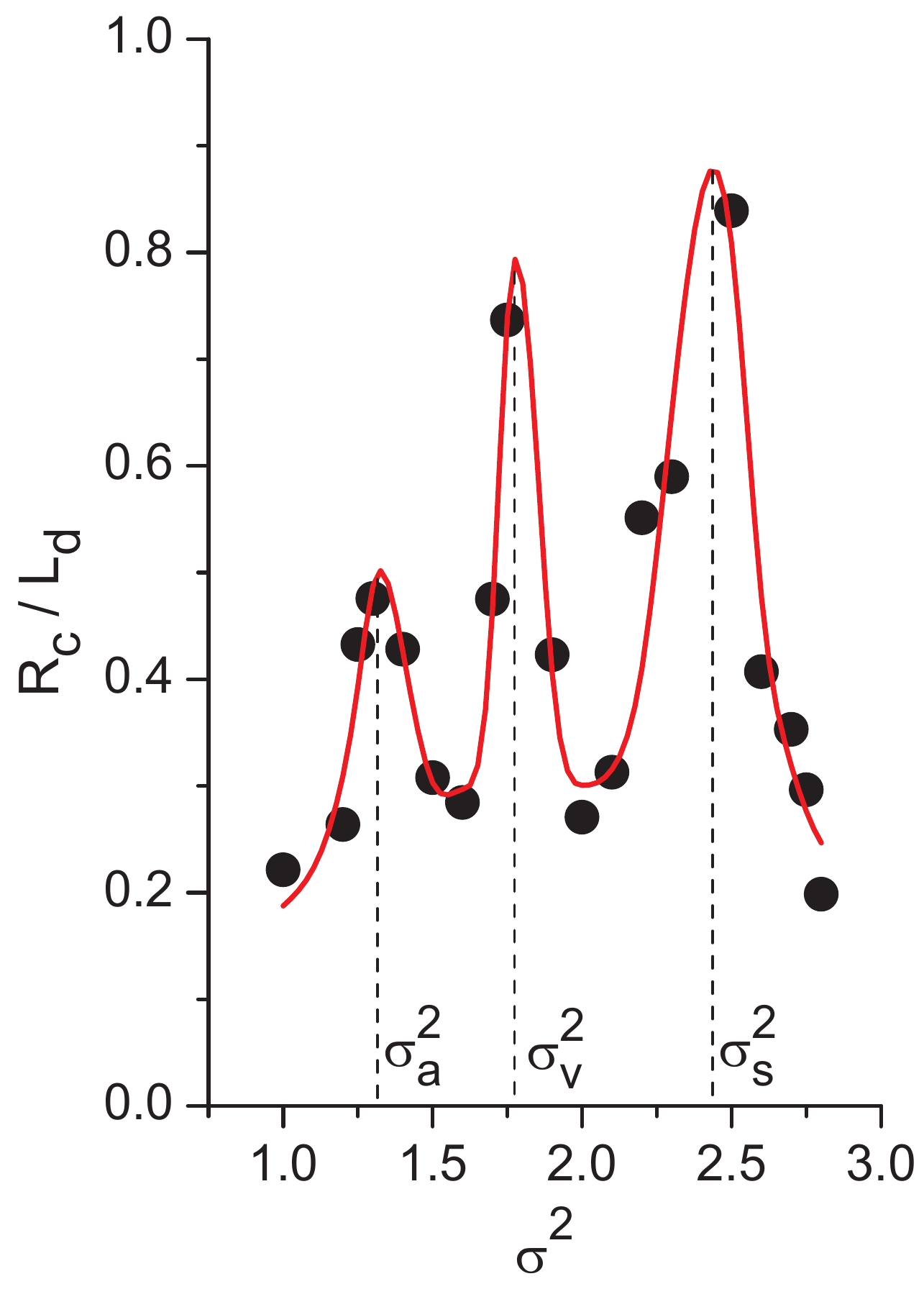}
c) \includegraphics[width=0.45\textwidth]{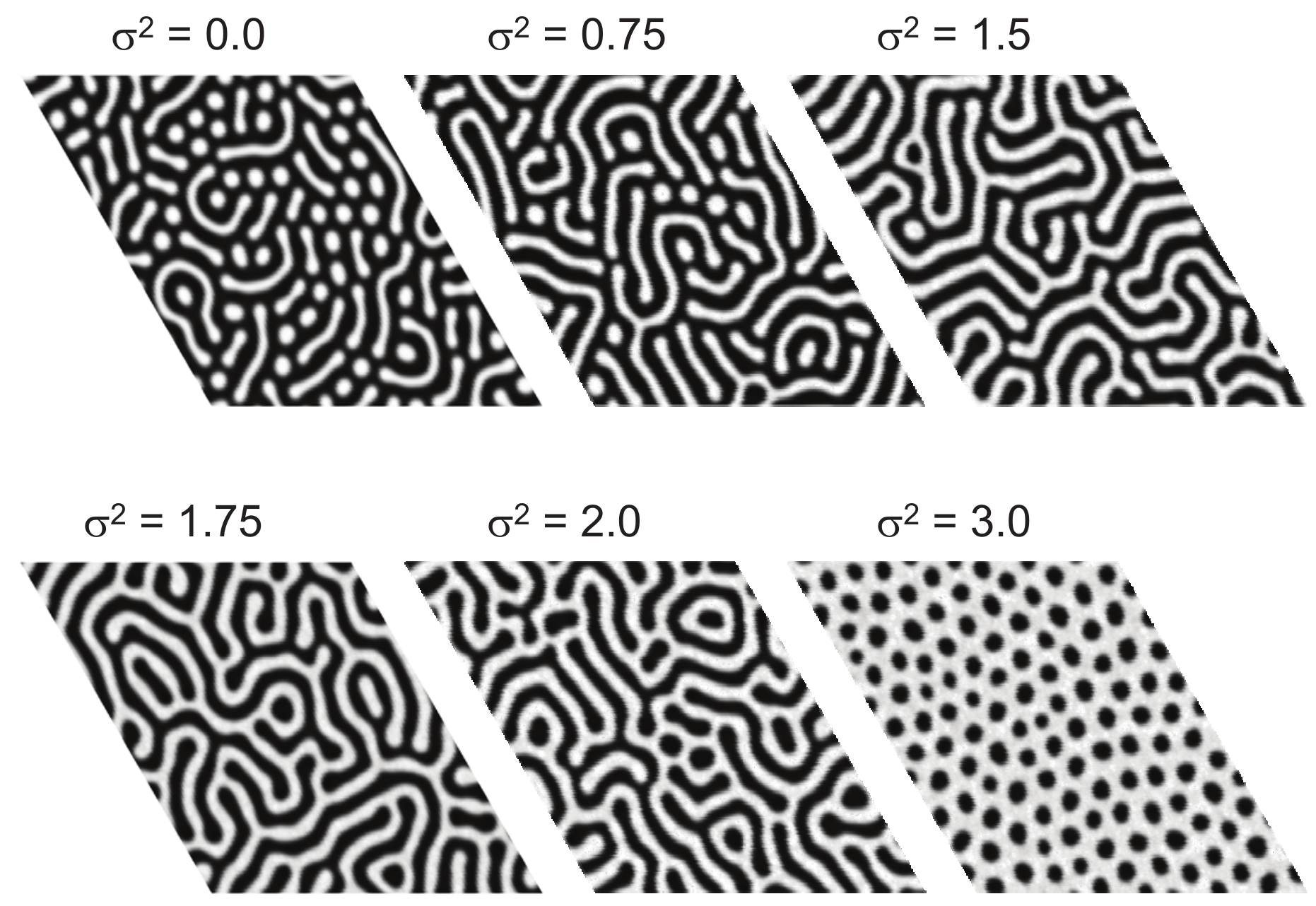}
\end{center}
\caption{(Colour online) Dependencies of the mean distance between structures $R_0$ (a) and 
correlation radius $R_c$ (b) on noise intensity $\sigma^2$ and (c) snapshots of the 
adsorbate configuration in the stationary limit at different~$\sigma^2$.  
Other parameters are: $\alpha=0.2$, $u_E=1.0$, $\varepsilon=3.5$.}
\label{fig8}
\end{figure}

The found transformation in the surface morphology can be effectively studied by considering the correlation properties 
of the coverage field $x({\bf r})$ in the quasi-stationary limit $t\to\infty$. To perform this analysis, 
we  consider the stationary two-point correlation function
$C(r)=\langle x(r)x(0)\rangle$, which can be represented in the form:
$C(r)=A\re^{-r/R_c}\cos(2\piup r/R_0+\phi)$. Here, $R_c$ and $R_0$ are the
correlation radius and the mean distance between structures (period of spatial 
modulations), respectively. Both quantities are sensitive to changes in morphology of the growing surface 
\cite{PRE12,NRL17,EPJB19}. The dependencies of the period of spatial modulations  
$R_0$ and the correlation radius $R_c$ \emph{versus} noise intensity $\sigma^2$ are shown 
in figures~\ref{fig8} a, b, respectively with the typical quasi-stationary snapshots in figure~\ref{fig8} c at 
$\alpha=0.2$, $u_E=1.0$ and $\varepsilon=3.5$.

It follows that with the fluctuation intensity growth, the period $R_0$ increases in a non-monotonous 
manner, having three peaks, located at $\sigma^2_a$, $\sigma^2_v$ and $\sigma^2_s$ (see figure~\ref{fig8}~a).
The dependence $R_c(\sigma^2)$ is also characterized by the three peaks with the same location. 
The last peak at $\sigma^2=\sigma^2_s$ is related to the maximal 
value of the stationary order parameter $\langle(\delta x)^2\rangle_\text{st}$ (compare with figure~\ref{fig7}~a).
By analysing the dependencies $R_0(\sigma^2)$ and $R_c(\sigma^2)$ with the snapshots in figure~\ref{fig8},
one can  argue that at $\sigma^2=\sigma^2_a$ the labyrinthine structure of percolating adsorbate 
clusters is realized (see snapshots at $\sigma^2=0.75$ and $\sigma^2=1.5$ in figure~\ref{fig8}~c), whereas at 
$\sigma^2=\sigma^2_v$, separated spherical holes inside the adsorbate matrix start to organize 
(see snapshots at $\sigma^2=1.75$ and $\sigma^2=2.0$ in figure~\ref{fig8}~c). The critical value $\sigma^2_a$ 
corresponding to the change in the surface morphology from the separated adsorbate islands 
on the substrate through the labyrinthine structure towards 
separated holes inside the adsorbate matrix relates well to $\sigma^2_m$, obtained in the framework of 
the stability analysis 
(see inset in figure~\ref{fig4}), when the stability exponent $\lambda(\kappa_m)$ has the maximal value. 

\begin{figure}
\begin{center}
a) \includegraphics[width=0.22\textwidth]{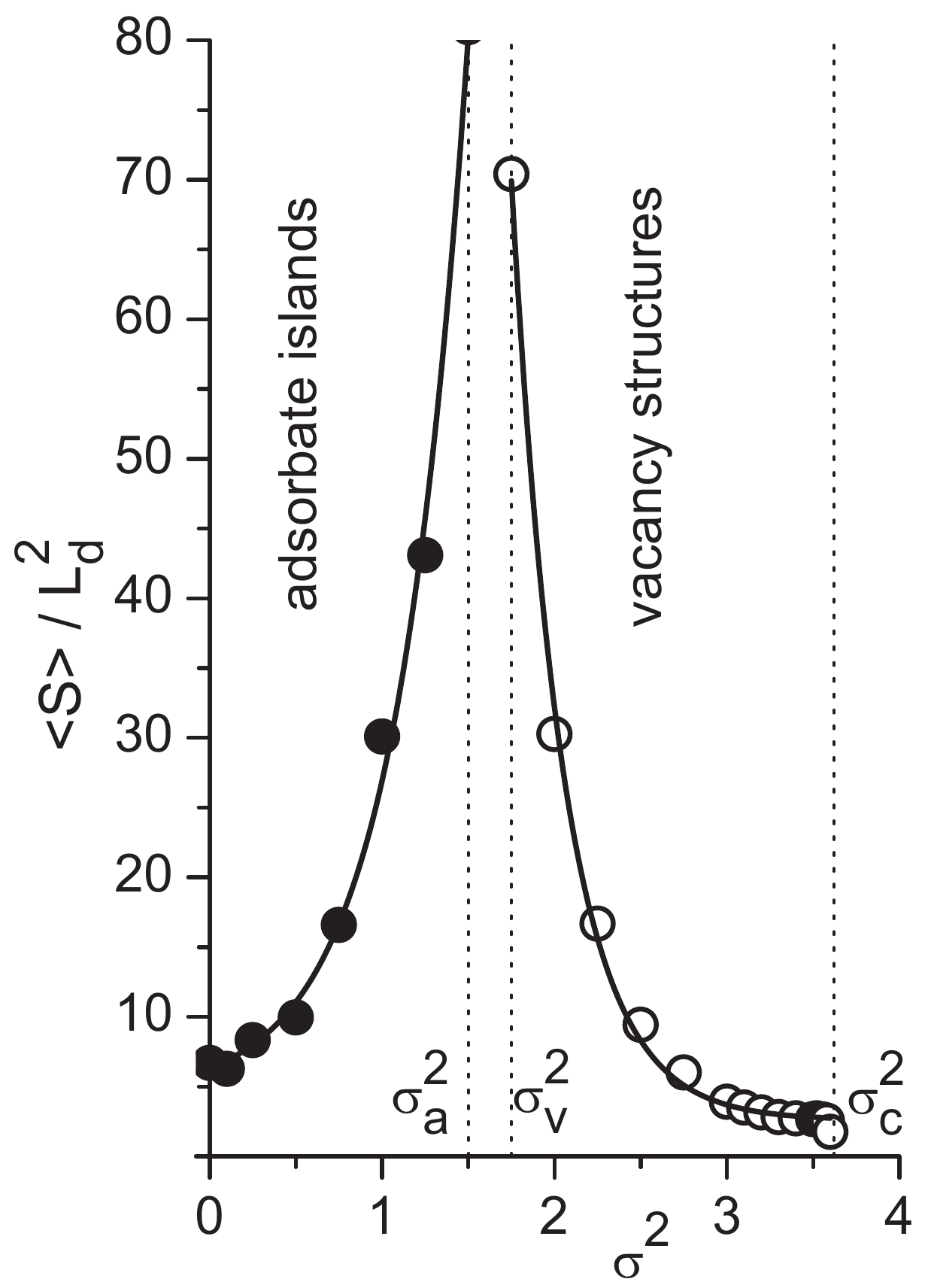}
b) \includegraphics[width=0.22\textwidth]{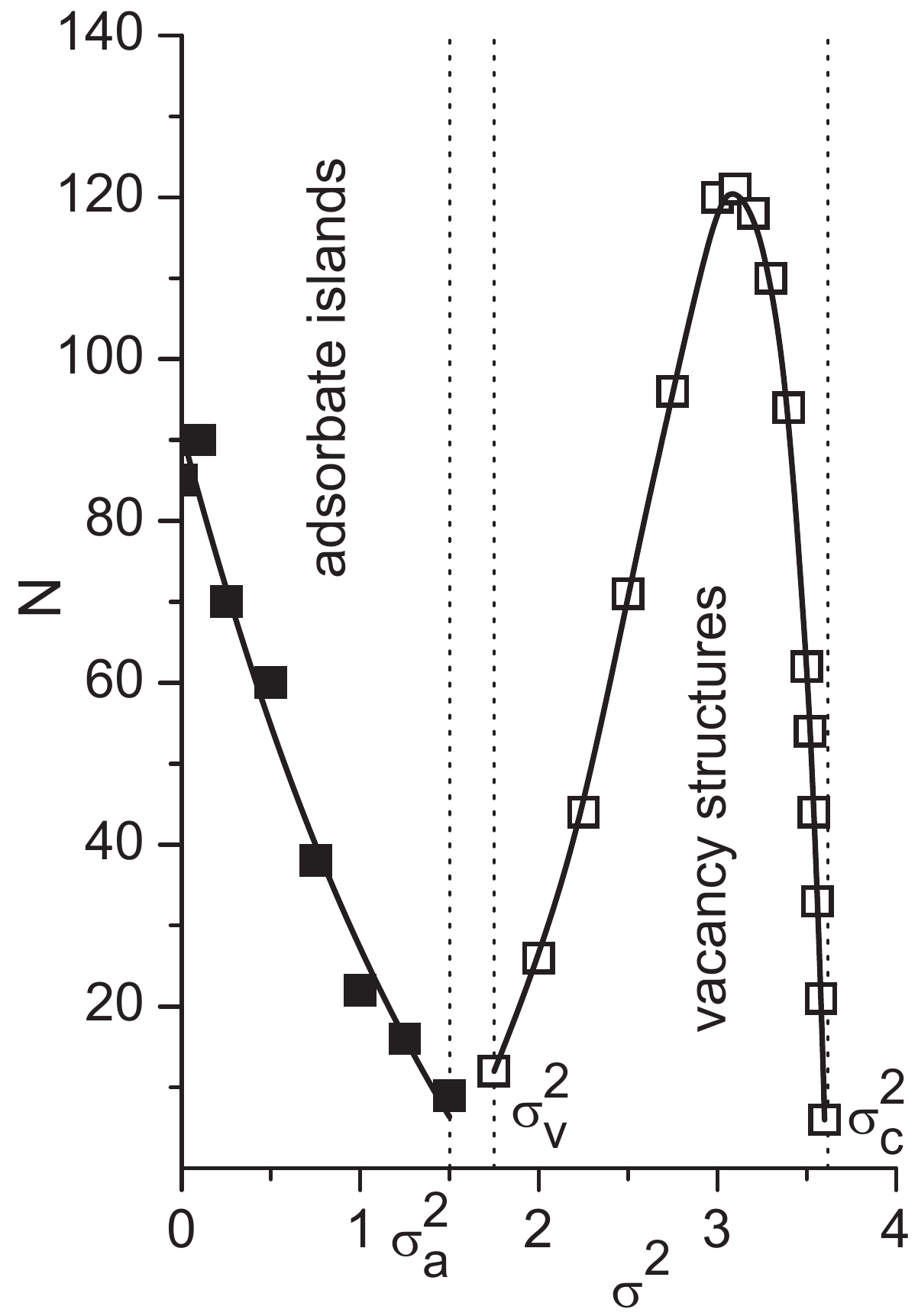}
\end{center}
\caption{Dependencies of the mean area of structures (a) and number of islands (b) on noise intensity~$\sigma^2$. Other parameters are: $\alpha=0.2$, $u_E=1.0$, $\varepsilon=3.5$.}
\label{fig9}
\end{figure}

To perform a statistical analysis of a change in the surface morphology in quasi-stationary limit 
with the noise intensity growth, we  calculated the mean area $\langle S\rangle$ of both adsorbate 
structures and vacancy structures (holes) and their number $N$ for different values of the noise intensity 
$\sigma^2$.  The obtained results are shown in figures~\ref{fig9}~a, b, respectively. 
First, let us consider the case of small values of the noise intensity $\sigma^2<\sigma^2_a$ when 
separated adsorbate islands are realized (see curves with filled circles and with filled squares in 
figures~\ref{fig9}~a, b, respectively). It follows that with the noise intensity growth, the mean area of the 
adsorbate island monotonously increases and their number drops. It means that fluctuations induce 
interactions between adsorbate islands, leading to the formation of elongated structures of adsorbate, i.e., transition 
to the labyrinthine-like pattern. Such type of the surface morphology can be observed if the fluctuation 
intensity $\sigma^2$ lies in the interval $(\sigma^2_a,\sigma^2_v)$. At $\sigma^2>\sigma^2_v$, separated 
holes inside the adsorbate matrix are realized. Their mean area decreases with the noise intensity growth 
meaning the formation of spherical-shaped vacancy structures (see empty circles in figure~\ref{fig9}~a). 
The number of vacancy islands grows with $\sigma^2$, attains maximal value and then drops to 
zero at $\sigma^2\to\sigma^2_c$ (see empty squares in figure~\ref{fig9}~b). A decrease in a number of 
$N$ of the vacancy structures relates with an increase in the stationary mean adsorbate concentration 
with the noise intensity growth (see dependence $\langle x \rangle_\text{st}$ in figure~\ref{fig7}~a).

 We have just studied the influence of the intensity of external field fluctuations onto the 
dynamics of pattern formation at fixed system parameters. At the same time, according to the constructed 
model of nano-structured thin films growth during deposition in plasma-condensate systems, the 
intensity $\sigma^2$ of fluctuations of the external (electrical) field is proportional to 
the mean value of this strength $u_E$. Hence, in real experiment, an increase in the mean strength of the 
external field results in the growth of  intensity of its fluctuations. The stability analysis provided 
in the previous section  and the performed  numerics show that an increase in both 
$u_E$ and $\sigma^2$ inside domain B in figure~\ref{fig3} provides the formation of stable surface structures 
during deposition. The found noise induced morphological transformation of the surface pattern from 
separated adsorbate structures towards separated holes indicates that separated compact nano-dots are 
realized at values of both $u_E$ and $\sigma^2$ near the top curve inside domain B (see left-hand 
panel in figure~\ref{fig3}). Next, we fix $\alpha=0.2$ and $\varepsilon=3.5$ and focus our attention 
on studying the influence of the noise-over-signal ratio (NOSr) $\sigma^2/u_E$ onto the change of the statistical 
properties of separated nano-dots by taking into account the functional dependence $\sigma^2=au_E-b$ 
with $u_E\geqslant2.5$ (see dot curve in left-hand panel in figure~\ref{fig3} with 
$a=6.18$ and $b=15.44$).   

\begin{figure}[!t]
\begin{center}
a)\includegraphics[width=0.45\textwidth]{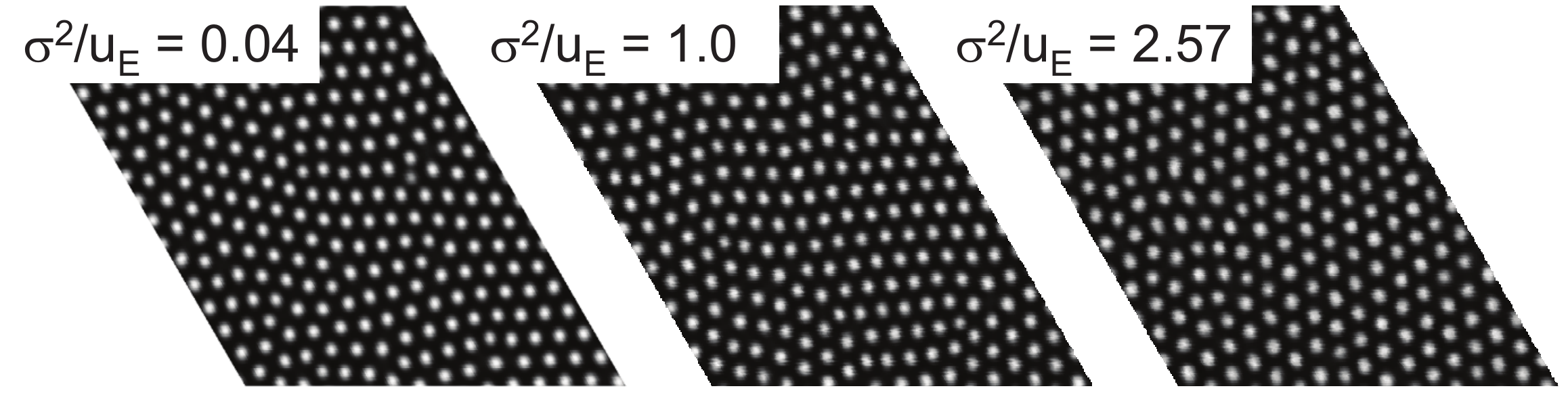}\\
b)\includegraphics[width=0.22\textwidth]{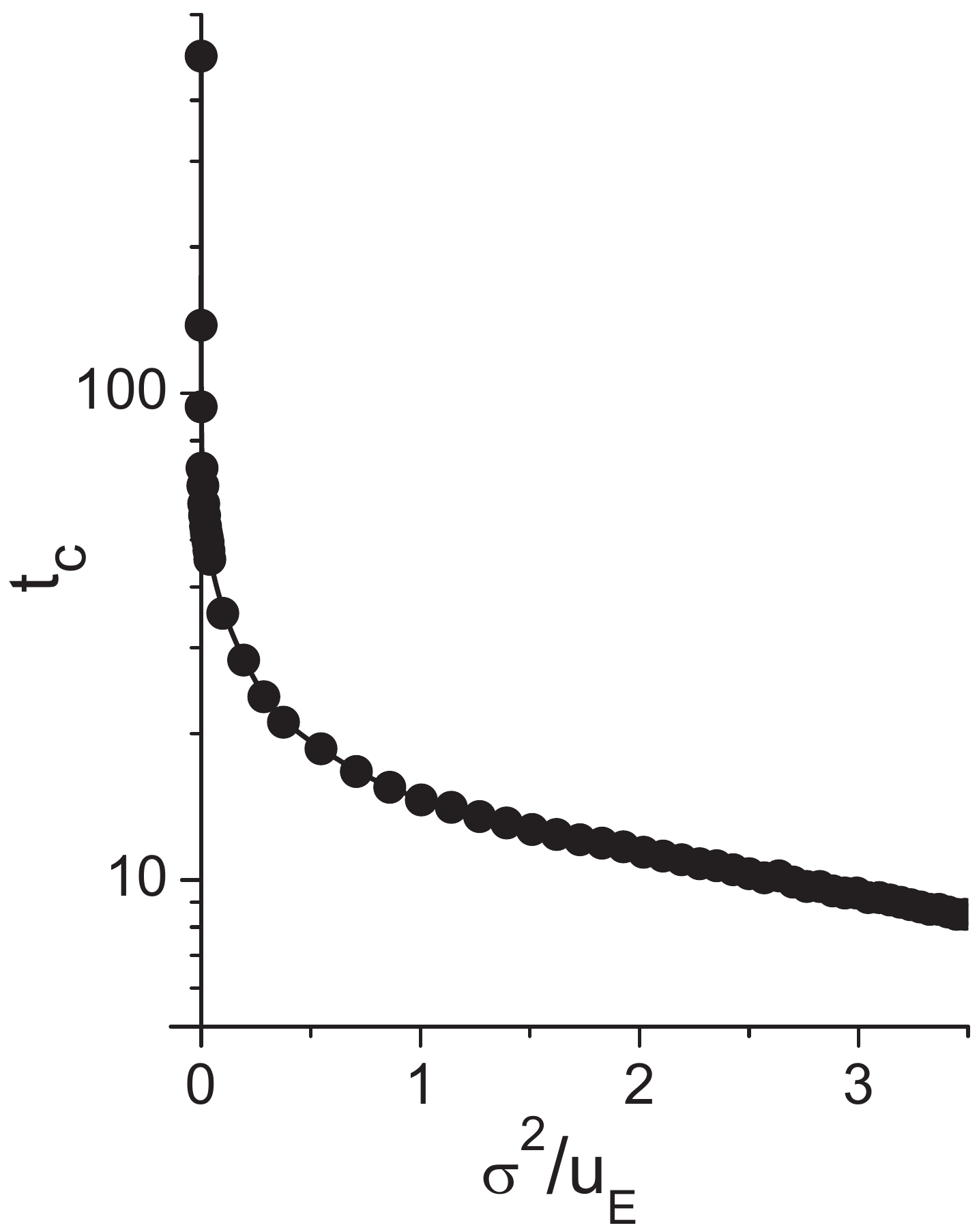}
c)\includegraphics[width=0.22\textwidth]{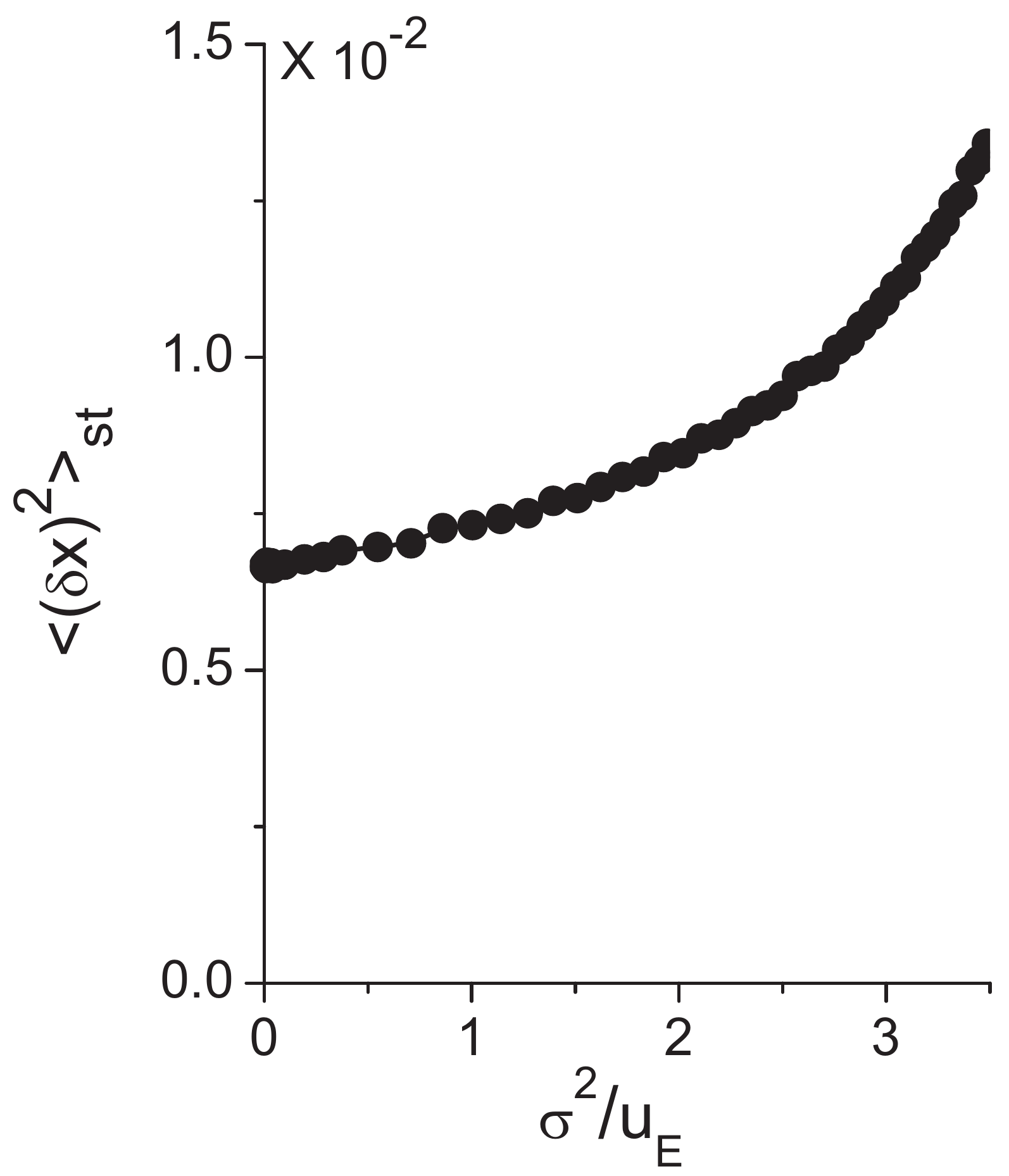}
\end{center}
\caption{Dependencies of: (a) the time instant $t_c$ when ordering starts; and (b) 
stationary value of the order parameter $\langle(\delta x)^2\rangle_\text{st}$
on the ratio $\sigma^2/u_E$.}
\label{fig10}
\end{figure}

First, we analyse the influence of the NOSr onto the dynamics of pattern formation 
and onto the order in spatial distribution of the coverage field in the plasma-condensate system studied. 
In figure~\ref{fig10}~a we show typical quasi-stationary snapshots of the surface morphology 
at different values of the NOSr $\sigma^2/u_E$. It follows that in the case of the functional dependence 
of the intensity $\sigma^2$ of fluctuations of the external flux on  its strength $u_E$, 
the morphology of the surface remains the same: separated adsorbate clusters. In figure~\ref{fig10}~b
we present the dependence of  the time instant $t_c$, indicating a start of formation of the islands 
 on the ratio $\sigma^2/u_E$. It follows that compared to the deterministic case with 
$\sigma^2/u_E=0$, an increase in the NOSr leads to an extreme decrease of the 
time instant $t_c$, meaning acceleration of the ordering processes. At large values of $\sigma^2/u_E$, 
time instant $t_c$ slightly decreases with the NOSr growth. A similar result was 
found at fixed $u_E$ by increasing the noise intensity $\sigma^2$ (see figure~\ref{fig6}). However, 
in that case such an acceleration was related to the fluctuations induced transformation of the 
surface morphology from separated adsorbate clusters towards separated holes inside the adsorbate matrix. 
In the actual case of  functional dependence of the noise intensity $\sigma^2$ on the mean value of  
anisotropy strength $u_E$, the morphology of the surface remains the same (see figure~\ref{fig10}~a). 
Hence, the acceleration of the ordering processes shown in figure~\ref{fig10}~b is related to an increasing 
influence of both deterministic and stochastic parts of the external flux, caused by the electrical 
field presence near the substrate.

In figure~\ref{fig10}~c we show the dependence of the stationary value of the order parameter 
$\langle (\delta x)^2\rangle$ on the NOSr $\sigma^2/u_E$. It follows that 
the order parameter continuously grows with $\sigma^2/u_E$ increasing, meaning realization of a 
more ordered surface. Again, this effect is caused by a simultaneous influence of anisotropy 
strength and its fluctuations, rather than the morphological transformation of the growing surface 
discussed before (see figures~\ref{fig7}~a, b).

\begin{figure}[!t]
\begin{center}
\includegraphics[width=0.40\textwidth]{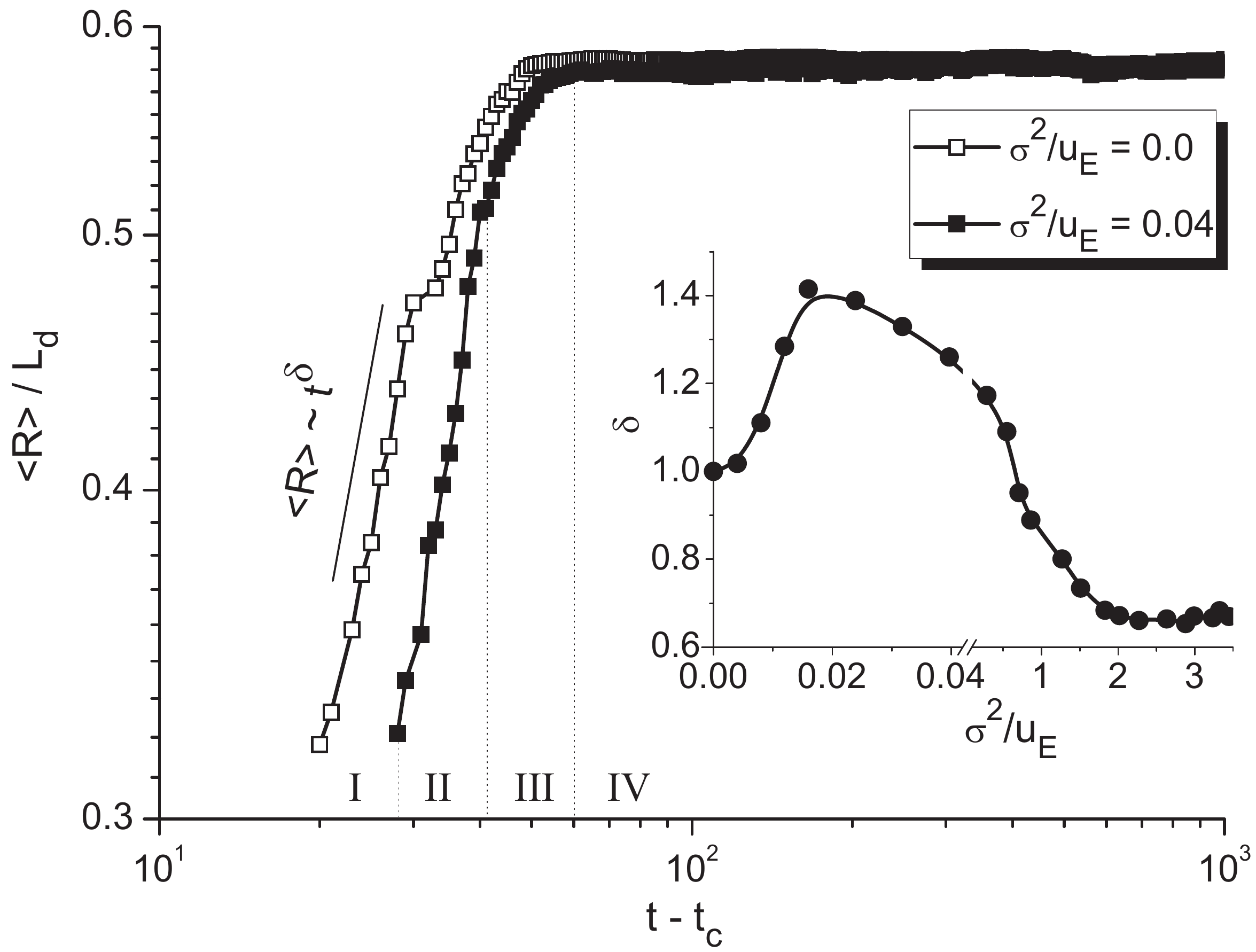}
\end{center}
\caption{Evolution of the mean linear size of adsorbate islands $\langle R\rangle$ in units of 
diffusion length, calculated after the incubation period $t_c$ at different values of the noise-over-signal 
ratio $\sigma^2/u_E$. Inset shows the dependence of the growth exponent $\delta$ on the ratio $\sigma^2/u_E$.}
\label{fig11}
\end{figure}

One of the intriguing problems in the formation of separated islands  (nucleation processes, grains and 
voids growth, etc.) is the growth law of the mean size of these structures. We have calculated the linear size of  each 
adsorbate cluster $R$  assuming $S=\piup R^2$. The dynamics of the averaged linear size 
$\langle R\rangle$ over all adsorbate structures for deterministic (empty squares) and stochastic 
(filled squares) systems is shown in figure~\ref{fig11}. 
Here, time is counted from the corresponding 
value $t_c$, which drops with ratio $\sigma^2/u_E$ (see figure~\ref{fig10}~b). 
One can issue four different stages of the mean radius $\langle R\rangle$ evolution 
(see curve with filled squares in figure~\ref{fig11}).
The first stage $I$ corresponds to the formation of adsorbate islands. At this stage, the concentration 
of adsorbate grows in time and interacting adatoms tend to organize separated adsorbate clusters. 
At the stage $II$, the formed small separated adsorbate clusters start to grow (growth stage). When 
growing stage is finished, the coarsening stage starts (stage $III$). Finally, the quasi-stationary 
regime of the mean size evolution is realized (stage $IV$). It follows that the growth stage is 
characterized by the power-law asymptotic $\langle R\rangle(t)\propto t^\delta$ with the growth 
exponent $\delta$. The dependence of the growth exponent $\delta$ on the 
NOSr $\sigma^2/u_E$ is shown in the inset in figure~\ref{fig11}. It follows that the 
NOSr affects the growth exponent. In the pure deterministic system at $\sigma^2=0$, 
one gets a normal law of the linear size of adsorbate islands. An increase of the ratio $\sigma^2/u_E$ 
provides an increase of the growth exponent up to $\delta\simeq1.4$ meaning the acceleration of the linear 
size growth. With a further growth in $\sigma^2/u_E$, the growth exponent $\delta$ drops, attaining 
quasi-stationary value $\delta\simeq0.65$ at $\sigma^2/u_E>2$ (see inset in figure~\ref{fig11}). 

\begin{figure}[!t]
\begin{center}
a)\includegraphics[width=0.22\textwidth]{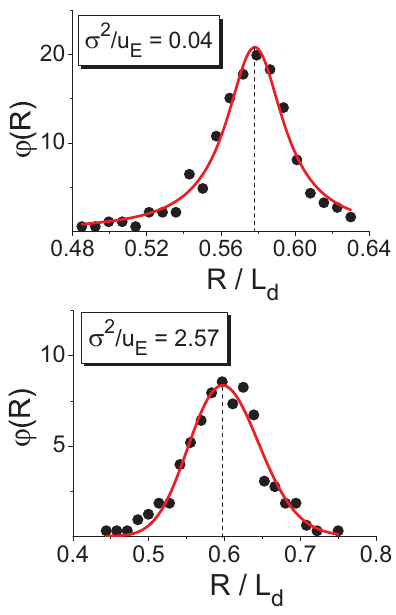}
b)\includegraphics[width=0.22\textwidth]{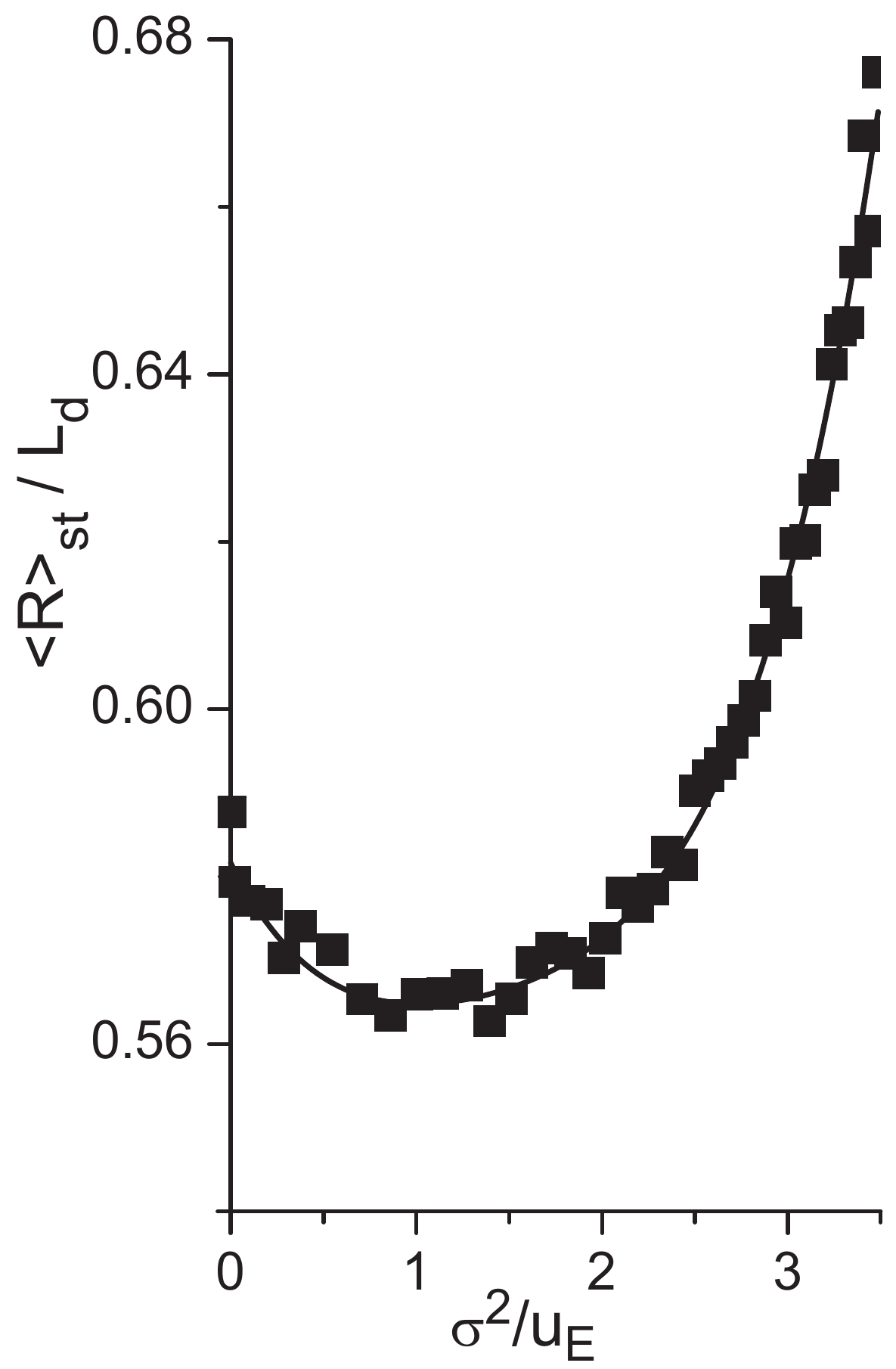}
\end{center}
\caption{(Colour online) Quasi-stationary snapshots and distributions of adsorbate islands over linear sizes in units of  diffusion length $L_\text{d}$ in quasi-stationary limit at different values of the 
the ratio $\sigma^2/u_E$ (a--c) and (d) the dependence of mean stationary value of the linear size of 
adsorbate islands $\langle R\rangle_\text{st}$ in units of diffusion length $L_\text{d}$ on the ratio $\sigma^2/u_E$.}
\label{fig12}
\end{figure}

Next, let us discuss the influence of the external flux caused by the electrical field onto the 
mean size of separated adsorbate islands in the quasi-stationary regime. The corresponding results are shown 
in figure~\ref{fig12}. Here, in figure~\ref{fig12}~a we show the distribution of adsorbate islands over sizes at small 
($\sigma^2/u_E=0.04$) and large ($\sigma^2/u_E=2.57$) values of the NOSr. Here, 
symbols correspond to the numerical data, which are fitted well by Lorenz distribution, shown by curves.
It is seen that an increase in the ratio $\sigma^2/u_E$ leads to the spreading of the distribution 
$\varphi(R)$, meaning that adsorbate islands are characterized by  different sizes. Moreover, the 
most probable value of the size of the adsorbate island, corresponding to the maximal value of the $\varphi(R)$ 
(shown by dash lines in figure~\ref{fig12}~a), increases with the ratio $\sigma^2/u_E$ growth. A detailed analysis 
of the influence of the ratio $\sigma^2/u_E$ onto the mean size of adsorbate islands allows us to obtain 
the dependence $\langle R\rangle_\text{st}(\sigma^2/u_E)$, shown in figure~\ref{fig12}~b. It is seen that 
with the growth of the NOSr, the quantity $\langle R\rangle_\text{st}$ decreases, attains 
the minimal value at $\sigma^2/u_E\simeq1$ and then monotonously increases. 

Hence, the combined influence of the deterministic and stochastic parts of the external flux 
at the regime of separated adsorbate islands formation controls the
dynamics of the pattern formation, the ordering of the growing surface, the growth law of the mean size of 
adsorbate islands and its stationary value.    

\section{Discussions}

\begin{figure}[!t]
\begin{center}
a)\includegraphics[width=0.45\columnwidth]{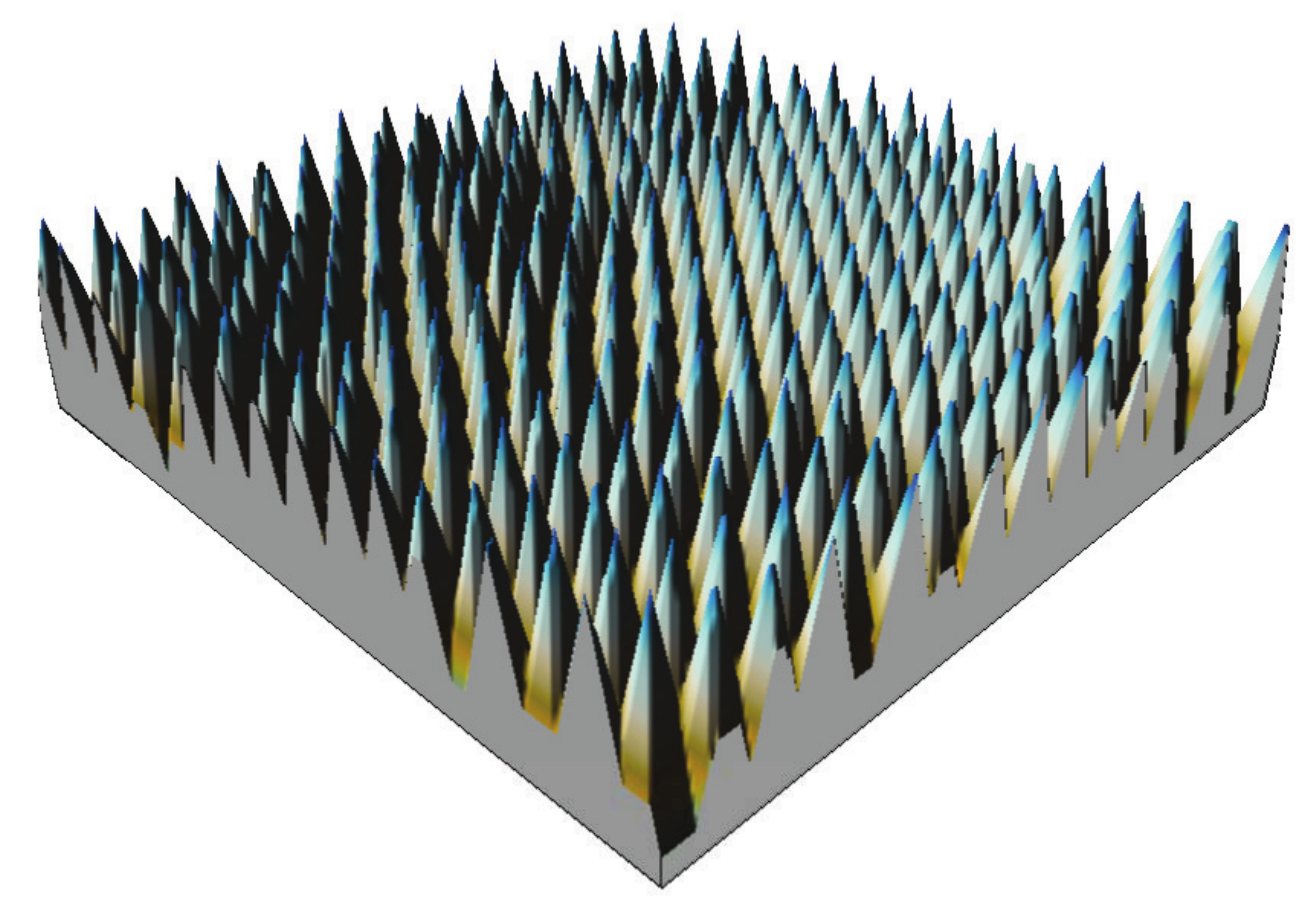}
b)\includegraphics[width=0.45\columnwidth]{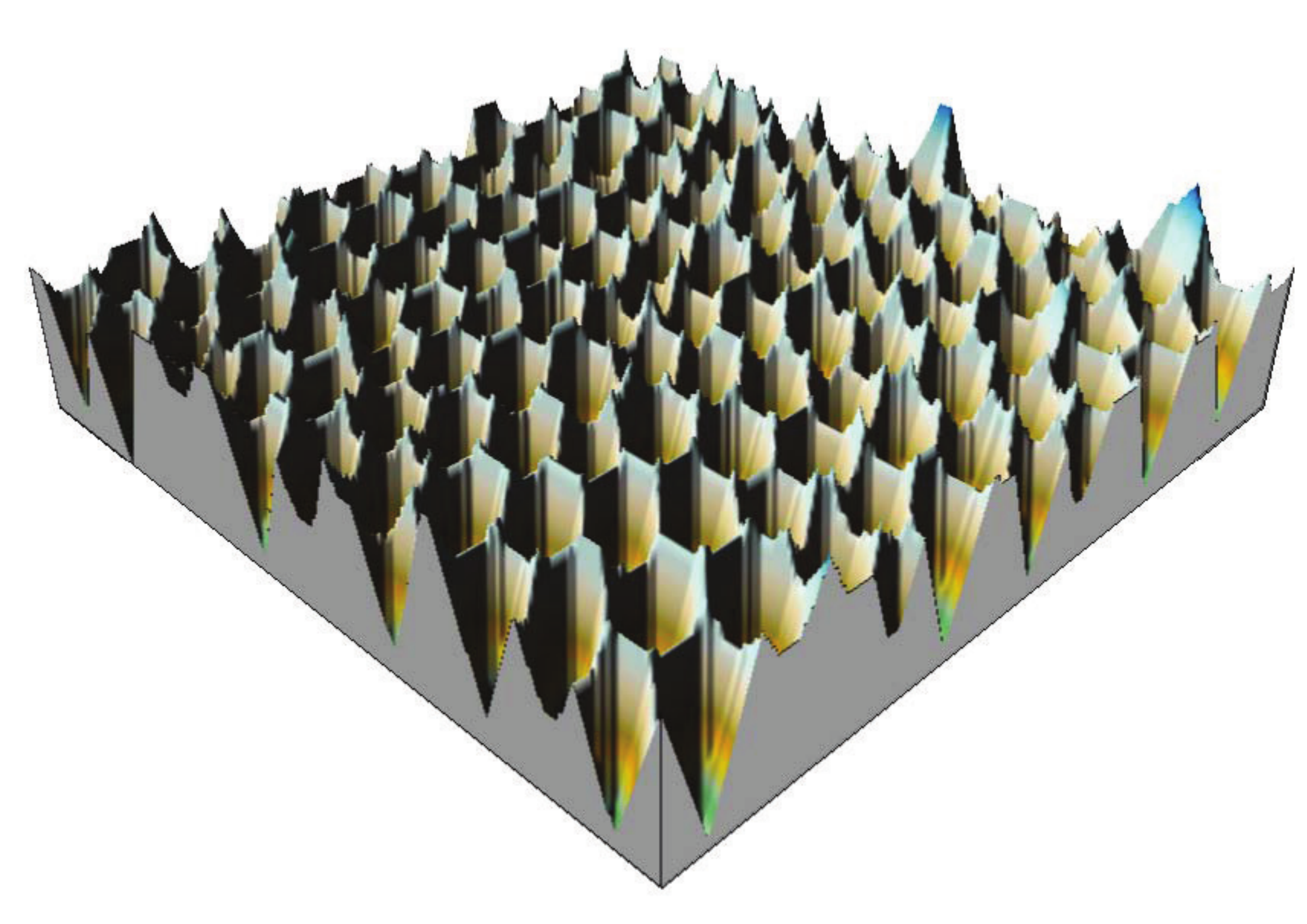}
\end{center}
\caption{(Colour online) Snapshots of the 3-d nano-dots (a) and nano-holes (b) with 16 and 21 layers, respectively, obtained at $\sigma^2=3$,     $\alpha=0.2$, $\varepsilon=3.5$ and  (a) $u_E=2.5$, (b) $u_E=1$.}
\label{fig13}
\end{figure}

The results presented  correspond to pattern formation on the intermediate $n$-th layer of 
the $N$-layers system. According to the 
constructed model, the terrace width of the multi-layer pyramidal-like structure constructed from all 
separated structures  is defined by the parameter $\Delta$, which for the actual case of computational  
grid linear size $L_0=256$ and fixed $\beta=0.1$ becomes $\Delta=13$ in units of the grid. 
To define the value of the terrace width $d$ for  each pyramidal-like multi-layer structure (see figure~\ref{fig1}) 
in units of the computational grid cites, we proceed in the following way.  
According to the definition of the total area, occupied by adsorbate on the $n$-th layer $S_n=\piup r_n^2$, 
for the corresponding total area on the precursor $(n-1)$-th layer, one gets $S_{n-1}=\piup r_{n-1}^2$, with 
$r_{n-1}=r_n+\Delta$. This yields:
\begin{equation}
S_{n-1}=S_n+2\Delta\sqrt{\piup S_n}+\piup\Delta^2. 
\label{snm11}
\end{equation}
On the other hand, the linear size of  each $i$-th pyramidal-like structure $r_{ni}$ decreases with 
the layer number $n$ growth by the terrace width $d$: $r_{n-1,i}=r_{n,i}+d$ (see figure~\ref{fig1}). 
Hence, the area of  each $i$-th structure on the $(n-1)$-th layer is: $s_{n-1,i}=s_{n,i}+2d\sqrt{\piup s_{n,i}}+\piup d^2$. 
Taking a sum over all $M$ structures,  one gets: 
\begin{equation}
S_{n-1}=S_{n}+2d\sum_i ^{M}\sqrt{\piup s_{ni}}+M\piup d^2. 
\label{snm12}
\end{equation}
By solving equations (\ref{snm11},\ref{snm12}) we get the terrace width $d$ for  each 
pyramidal-like structure as a solution of the quadratic equation in the form:
\begin{equation}
d=
\frac{1}{\piup M}
\left\{
\left[
\left(
\sum\limits_{i=1}^M\sqrt{\piup s_{ni}}
\right)^2
+
\piup M 
\left(
2\Delta\sqrt{\piup S_n}+\piup\Delta^2
\right)
\right]^{1/2}-\sum\limits_{i=1}^M\sqrt{\piup s_{ni}}
\right\}.
\label{eq4d}
\end{equation}
Hence, by using the calculated area of  each structure $s_{ni}$ and total number of structures $M$, we can get the size 
of the terrace width $d$ in units of the size of the computational grid $L_0$ and construct a multi-layer
system of $N$ layers, where the linear size of  each $i$-th structure decreases by $d$ with the layer number $n$ growth. 
For the actual case of $L_0=256$ and $\beta=0.1$ from equation~(\ref{eq4d}), we get $d=1$.
We take into account that the first layer is fully occupied by adsorbate 
and the highest layer is characterized by adsorbate structures with linear size, which equals  the terrace width  $d$.
Results for the constructed 3-dimensional $N$-layer structures with the terrace width $d=1$ at different values of the anisotropy strength $u_E$ are shown in figure~\ref{fig13}.

To provide the estimation of the obtained data for the linear size of the separated holes and 
adsorbate structures, we exploit the formula for the diffusion length
$L_\text{d}=a\exp([E_\text{d}-E_D]/2T)$ and use the typical data for semiconductors (Ge/Si(100)): 
the activation energy of adatom formation $E_\text{a}=0.6$~eV; the activation energy for desorption $E_\text{d}=1.25$~eV, the activation energy for diffusion $E_D=0.65$~eV and the lattice oscillation frequency $\nu=10^{12}$~$s^{-1}$. By taking lattice constant $a=5.6\times10^{-10}$~m for Ge and adatoms interaction strength $\epsilon=0.27$~eV, for the actual values of the temperature inside the chamber $T=773$~K, one gets $L_\text{d}\simeq50\times10^{-8}$~m. Hence, for the linear size of the unit cell, one gets $\ell\simeq8$ nm and for the terrace width of each pyramidal structure, one gets $d=\ell\simeq8$~nm which is in good agreement with the experimental results \cite{terrace1,terrace2,terrace3,terrace4}.
Therefore, from the data for  linear size of the spherically shaped adsorbate islands  
shown in figure~\ref{fig12}~b, one has $R\in(28\div35)$~nm  on 
the approximately half-height of the multi-layer system. These results correspond well with experimentally observed  data for nano-structured thin films growth at condensation \cite{bibitem52,bibitem54,bibitem50,bibitem51}.

\section{Conclusions}

In this article we have studied the combined influence of both deterministic and stochastic 
parts of the external flux onto the dynamics and statistical properties of pattern formation in 
plasma-condensate devices. By summarizing all obtained results, one can highlight the main conclusions.

1. Fluctuations of the strength of the external field induce first-order phase transition in a
homogeneous system. 

2. At fixed values of the mean intensity of the strength of electrical field near the substrate, 
the fluctuations of strength induce the ordering of the adsorbate on the substrate, leading to the formation 
of separated nano-sized adsorbate islands. An increase in the fluctuation intensity leads to:
(i) a morphological transformation of the surface from separated adsorbate islands through the percolating 
structure of adsorbate towards separated nano-holes inside the adsorbate matrix; (ii) an acceleration of the 
ordering processes; (iii) an increase in the spatial order of the growing surface.   
Strong fluctuations stabilize the system resulting in homogenization of the coverage field. 

3. In the case of  correlations between the mean value of the strength of the electrical field and 
its fluctuation intensity in the regime of nano-dots formation, an increase in the noise-over-signal 
ratio results in acceleration of the pattern formation; provides the formation of well ordered patterns; 
leads to anomalous dynamics of the mean size of adsorbate islands. The stationary value of the 
mean adsorbate islands can be minimized with variation in the noise-over-signal ratio and           
varies in the interval $R\in(28,35)$~nm for the prototype systems: semiconductor on silicate.

The derived effective one-layer model can be used to modelize nano-structured thin films 
growth at deposition in 3-dimensional space with the formation of 
separated cone-like adsorbate structures and smooth multi-layer holes inside the adsorbate matrix.
 
The results obtained within this work  extend the existing knowledge 
about controlling the dynamics of pattern formation and statistical properties of surface structures and 
show a constructive role of the external fluctuations, which is capable of controlling the patterning and the scaling 
behaviour of the system. 

We expect that our non-trivial findings will stimulate further theoretical and experimental 
studies in the field of nano-structured thin film growth in plasma-condensate devices.   

\section*{Acknowledgements}

Support of this research by the Ministry of Education and Science of Ukraine, 
project No. 0117U003927, is gratefully acknowledged.

%

\newpage
\ukrainianpart

\title{Індуковані шумом ефекти у процесах росту наноструктурованих тонких плівок при осадженні в системах плазма-конденсат}%
\author{В.О. Харченко\refaddr{label1,label2}, А.В. Дворниченко\refaddr{label2}, Д.О. Харченко\refaddr{label1}}
\addresses{
\addr{label1}Інститут прикладної фізики НАН України, вул. Петропавлівська 58, 40000 Суми, Україна
\addr{label2}Сумський державний університет, вул. Римського-Корсакова 2, 40007 Суми, Україна 
}
\makeukrtitle
\begin{abstract}
\tolerance=3000%

У даній роботі проводиться всебічне дослідження індукованих шумом ефектів у 
стохастичній моделі реакційно-дифузійного типу, що описує процес зростання 
наноструктурованих тонких плівок при конденсації в системі плазма-конденсат.
Вводиться зовнішній потік адсорбату між сусідніми шарами, спричинений 
наявністю електричного поля біля підкладки. Враховується, що напруженість
електричного поля флуктуює навколо його середнього значення.
Обговорюється конкуруючий вплив регулярної та стохастичної частин зовнішнього 
потоку на динаміку системи. Показано, що введені зовнішні флуктуації
здатні індукувати фазовий перехід першого порядку в однорідній системі, 
керувати формуванням структур у просторово-розподіленій системі. Такі флуктуації
контролюють динаміку формування поверхневих структур, просторовий порядок, 
морфологію поверхні, закон зростання середнього розміру островів адсорбату, 
тип та лінійний розмір поверхневих структур. Детально проаналізовано вплив 
інтенсивності введених флуктуацій на скейлінгові та статистичні властивості
наноструктурованої поверхні. Отримані результати забезпечують розуміння
деталей індукованих шумом ефектів при формуванні поверхневих нанорозмірних 
структур у процесах конденсації в системах плазма-конденсат.

\keywords стохастичні системи, нелінійна динаміка, структуроутворення, індуковані шумом ефекти  
 
\end{abstract}


\begin{thebibliography}{00}
\bibitem{poster1} Cross~M.C., Hohenberg~P.C., Rev. Mod. Phys., 1993, \textbf{65}, 851, \doi{10.1103/RevModPhys.65.851}.
\bibitem{NIE1} Sharma~A., Khanna~R., Phys. Rev. Lett., 1998, \textbf{81}, 3463, \doi{10.1103/PhysRevLett.81.3463}.
\bibitem{NIE2} Hansen~ J.L., van Hecke~M., Haaning~A., Ellegaard C., Andersen K.H., Bohr T., Sams T., Nature, 2001, \textbf{410}, 324, \doi{10.1038/35066631}.
\bibitem{NIE3} Elbelrhiti~H., Claudin~P., Andreotti~B., Nature, 2005, \textbf{437}, 720, \doi{10.1038/nature04058}.
\bibitem{NIE4} Langer~J.S., Rev. Mod. Phys., 1980, \textbf{52}, 1, \doi{10.1103/RevModPhys.52.1}.
\bibitem{NIE5} Onorato~M., Osborne~A.R., Serio~M., Bertone~S., Phys. Rev. Lett., 2001, \textbf{86}, 5831, \\ \doi{10.1103/PhysRevLett.86.5831}.
\bibitem{NIE6} Treiber~M., Kramer~L., Phys. Rev. E, 1994, \textbf{49}, 3184, \doi{10.1103/PhysRevE.49.3184}.
\bibitem{NIE8} Jung~P., Mayer-Kress~G., Phys. Rev. Lett., 1995, \textbf{74}, 2130, \doi{10.1103/PhysRevLett.74.2130}.
\bibitem{NIE9} Wang~J., Kadar~S., Jung~P., Showalter~K., Phys. Rev. Lett., 1999, \textbf{82}, 855, \doi{10.1103/PhysRevLett.82.855}.
\bibitem{NIE10} Hou~Z., Xin~H., Phys. Rev. Lett., 2002, \textbf{89}, 280601, \doi{10.1103/PhysRevLett.89.280601}.
\bibitem{NIE11} Hempel~H., Schimansky-Geier~L., Garc\'ia-Ojalvo~J., Phys. Rev. Lett., 1999, \textbf{82}, 3713, \\ \doi{10.1103/PhysRevLett.82.3713}.
\bibitem{NIE14} Biancalani~T., Dyson~L., McKane~A.J., Phys. Rev. Lett., 2014,  \textbf{112}, 038101, \\ \doi{10.1103/PhysRevLett.112.038101}.
\bibitem{NIE16} Weiss~T., Kronwald~A., Marquardt~F., New J. Phys., 2016, \textbf{18}, 1, \doi{10.1088/1367-2630/18/1/013043}.
\bibitem{PhysScr11} Kharchenko~D.O., Kharchenko~V.O., Lysenko~I.O., Physica Scripta, 2011, \textbf{83}, 045802, \\ \doi{10.1088/0031-8949/83/04/045802}.
\bibitem{JPS1} Horsthemke~W., Lefever~R., Noise-Induced Transitions, Springer-Verlag, Berlin, 1984.
\bibitem{JPS2} Garcia-Ojalvo~J., Sancho~J.M., Noise in Spatially Extended System, Springer-Verlag, New York, 1999.
\bibitem{JPS4} Van den Broeck~C., Parrondo~J.M.R., Toral~R., Phys. Rev. Lett., 1994,  \textbf{73}, 3395, \\\doi{10.1103/PhysRevLett.73.3395}.
\bibitem{JPS5} Van den Broeck~C., Parrondo~J.M.R., Toral~R., Kawai~R., Phys. Rev. E, 1997, \textbf{55}, 4084, \\ \doi{10.1103/PhysRevE.55.4084}.
\bibitem{Elder} Elder~K.R., Grant~M., Phys. Rev. E., 2004, \textbf{70}, 051605, \doi{10.1103/PhysRevE.70.051605}.
\bibitem{PhysA2010} Kharchenko~D., Lysenko~I., Kharchenko~V., Physica A, 2010, \textbf{389}, 3356, \doi{10.1016/j.physa.2010.04.027}.
\bibitem{Buceta9} Garcia-Ojalvo~J., Hernandez-Machado~A., Sancho~J.M., Phys. Rev. Lett., 1993, \textbf{71}, 1542, \\ \doi{10.1103/PhysRevLett.71.1542}.
\bibitem{Buceta10} Becker~A., Kramer~L., Phys. Rev. Lett., 1994, \textbf{73}, 955, \doi{10.1103/PhysRevLett.73.955}.
\bibitem{Buceta11} Parrondo~J.M.R., Van den Broeck~C., Buceta~J., De la Rubia F.J., Physica A, 1996, \textbf{224}, 153,\\ \doi{10.1016/0378-4371(95)00350-9}.
\bibitem{Buceta12} Zaikin~A.A., Schimansky-Geier~L., Phys. Rev. E, 1998, \textbf{58}, 4355, \doi{10.1103/PhysRevE.58.4355}.
\bibitem{Castets1990} Castets~V.V., Dulos~E., Boissonade~J., de Kepper~P., Phys. Rev. Lett., 1990, \textbf{64}, 2953, \\ \doi{10.1103/PhysRevLett.64.2953}.
\bibitem{RDS1} Nicolis~G., Prigogine~I., Self-Organization in Nonequilibrium Systems, Wiley, New York, 1977.
\bibitem{RDS2} Kuramoto~Y., Chemical Oscillations, Waves and Turbulence, Springer, Berlin, 1984.
\bibitem{RDS3} Epstein~I., Pojman~J., An Introduction to Nonlinear Chemical Dynamics, Oxford University press, Oxford, 1998.
\bibitem{PhysScr12} Kharchenko~V.O., Kharchenko~D.O., Kokhan~S.V.,  Vernyhora I.V.,   Yanovsky V.V., Phys. Scr., 2012, \textbf{86}, 055401, \doi{10.1088/0031-8949/86/05/055401}.
\bibitem{Mikhailov1} Hildebrand~M., Mikhailov~A.S., Ertl~G., Phys. Rev. Lett., 1998,  \textbf{81}, 2602, \doi{10.1103/PhysRevLett.81.2602}.
\bibitem{Mikhailov2} Hildebrand~M., Mikhailov~A.S., Ertl~G., Phys. Rev. E, 1998,  \textbf{58}, 5483, \doi{10.1103/PhysRevE.58.5483}.
\bibitem{PRE12} Kharchenko~V.O., Kharchenko~D.O., Phys. Rev. E, 2012, \textbf{86}, 041143, \doi{10.1103/PhysRevE.86.041143}.
\bibitem{CWM2002} Casal S.B., Wio H.S., Mangioni S., Physica A, 2002, \textbf{311}, 443, \doi{10.1016/S0378-4371(02)00828-2}.
\bibitem{NRL17} Kharchenko~V.O., Kharchenko~D.O., Yanovsky~V.V., Nanoscale Res. Lett., 2017, \textbf{12}, 337, \\ \doi{10.1186/s11671-017-2096-7}.
\bibitem{SS15} Kharchenko~V.O., Kharchenko~D.O., Surf. Sci., 2015, \textbf{637--638}, 90, \doi{10.1016/j.susc.2015.03.025}.
\bibitem{Wolgraef2003} Walgraef~D., Physica E, 2003, \textbf{18}, 393, \doi{10.1016/S1386-9477(02)01104-9}.
\bibitem{Wolgraef2004} Walgraef~D., Int. J. Quant. Chem., 2004, \textbf{98}, 248, \doi{10.1002/qua.10877}.
\bibitem{SS14} Kharchenko~V.O., Kharchenko~D.O., Dvornichenko~A.V., Surface Science, 2014, \textbf{630}, 158,  \\ \doi{10.1016/j.susc.2014.08.008}.
\bibitem{Walg18} Mishin Y., Farkas D., Mehl M.J., Papaconstantopoulos D.A., Phys. Rev. B,  1999, \textbf{59}, 3393, \\ \doi{10.1103/PhysRevB.59.3393}.
\bibitem{CrysGrowth19} Dvornichenko~A.V., Kharchenko~D.O., Lysenko~I.O., Kharchenko~V.O., J. Cryst. Growth, 2019, \textbf{514}, 1, \doi{10.1016/j.jcrysgro.2019.02.048}.
\bibitem{JAC} Wei~S., Ma~H.C., Chen~J.Q., Guo~J.D., J. Alloys Compd., 2016, \textbf{687}, 999, \doi{10.1016/j.jallcom.2016.06.253}.
\bibitem{spectr} Oh~Y.-H., Kim~S.-I., Kim~M., Lee~S.-Y., Kim~Y-W., Ultramicroscopy, 2017, \textbf{181}, 160, \\ \doi{10.1016/j.ultramic.2017.05.018}.
\bibitem{Perekrestov1} Perekrestov~V.I., Olemskoi~A.I., Kosminska~Yu.O., Mokrenko~A.A., Phys. Lett. A, 2009, \textbf{373}, 3386,\\ \doi{10.1016/j.physleta.2009.07.032}.
\bibitem{Perekrestov2} Kosminska~Y.A., Mokrenko~A.A., Perekrestov~V.I., Tech. Phys. Lett., 2011, \textbf{37}, 538, \\ \doi{10.1134/S1063785011060083}.
\bibitem{terrace1} Miccio~L.A., Setvin~M., M\"uller~M., Abad\'ia~M., Piquero~I., Lobo-Checa J., Schiller F., Rogero C., Schmid M., S\'anchez-Portal D., Diebold U., Ortega J.I.,  Nano Lett., 2016, \textbf{16}, 2017, \doi{10.1021/acs.nanolett.5b05286}.
\bibitem{terrace2} Ortega~J., Corso~M., Abd El-Fattah~Z.,  Goiri E.A., Schiller F., Phys. Rev. B, 2011, \textbf{83}, 085411,\\ \doi{10.1103/PhysRevB.83.085411}. 
\bibitem{terrace3} Neel~N., Maroutian~T., Douillard~L., Ernst~H-J., Phys. Rev. Lett., 2003, \textbf{91}, 226103, \\ \doi{10.1103/PhysRevLett.91.226103}. 
\bibitem{terrace4}  Czubanowski M., Schuster A., Akbari S., Pfn\"ur H.,  Tegenkamp C.,  New J. Phys., 2007, \textbf{9}, 338,\\ \doi{10.1088/1367-2630/9/9/338}.
\bibitem{VanKampen} Van Kampen~N.G., Stochastic Processes in Physics and Chemistry, North–Holland, Amsterdam, 1992.
\bibitem{EPJB18} Kharchenko~V.O., Dvornichenko~A.V., Borysiuk~V.N., Eur. Phys. J. B, 2018, \textbf{91}, 93, \\ \doi{10.1140/epjb/e2018-80730-8}.
\bibitem{PRE2010cite31} Sancho~J.M., San Miguel~M., Katz~S.L., Gunton~J.D., Phys. Rev. A, 1982, \textbf{26}, 1589, \\ \doi{10.1103/PhysRevA.26.1589}.
\bibitem{PRE2010cite32} Box~G.E.P., Muller~M.E., Ann. Math. Stat., 1958, \textbf{29}, 610, \doi{10.1214/aoms/1177706645}.
\bibitem{EPJB19} Kharchenko~V.O., Dvornichenko~A.V., Eur. Phys. J. B, 2019, \textbf{92}, 57, \doi{10.1140/epjb/e2019-90588-9}.
\bibitem{bibitem52} Leonhardt~D., Han~S.M., Surface Science, 2009, \textbf{603}, 2624, \doi{10.1016/j.susc.2009.06.015}.

\bibitem{bibitem54} Hunter~K.I., Held~J.T., Mkhoyan~K.A., Kortshagen~U.R., ACS Appl. Mater. Interfaces, 2017, \textbf{9}, 8263,\\ \doi{10.1021/acsami.6b16170}.
\bibitem{bibitem50} Perekrestov~V.I., Kosminska~Yu.O., Kornyushchenko~A.S., Latyshev~V.M., Physica B, 2013, \textbf{411}, 140,\\ \doi{10.1016/j.physb.2012.11.036}.
\bibitem{bibitem51} Kornyushchenko~A.S., Natalich~V.V., Perekrestov~V.I., J. Cryst. Growth, 2016, \textbf{442}, 68,\\ \doi{10.1016/j.jcrysgro.2016.02.033}.
\end{thebibliography}
\end{document}